\documentstyle[12pt,epsf,world_sci]{article}

\newlength{\extraspace}
\newlength{\extraspaces}
\newcommand{\be}{\begin{equation}
\addtolength{\abovedisplayskip}{\extraspaces}
\addtolength{\belowdisplayskip}{\extraspaces}
\addtolength{\abovedisplayshortskip}{\extraspace}
\addtolength{\belowdisplayshortskip}{\extraspace}}
\newcommand{\ee}{\end{equation}}
\newcommand{\bea}{\begin{eqnarray}
\addtolength{\abovedisplayskip}{\extraspaces}
\addtolength{\belowdisplayskip}{\extraspaces}
\addtolength{\abovedisplayshortskip}{\extraspace}
\addtolength{\belowdisplayshortskip}{\extraspace}}
\newcommand{\eea}{\end{eqnarray}}
\def\simge{\mathrel{%
   \rlap{\raise 0.511ex \hbox{$>$}}{\lower 0.511ex \hbox{$\sim$}}}}
\def\simle{\mathrel{
   \rlap{\raise 0.511ex \hbox{$<$}}{\lower 0.511ex \hbox{$\sim$}}}}
\def\slashchar#1{\setbox0=\hbox{$#1$}           
   \dimen0=\wd0                                 
   \setbox1=\hbox{/} \dimen1=\wd1               
   \ifdim\dimen0>\dimen1                        
      \rlap{\hbox to \dimen0{\hfil/\hfil}}      
      #1                                        
   \else                                        
      \rlap{\hbox to \dimen1{\hfil$#1$\hfil}}   
      /                                         
   \fi}                                         %

\newcounter{tabnum}
\newcommand{\D}{{\Delta}}

\newcommand{\g}{{\gamma}}

\renewcommand{\l}{{\lambda}}
\renewcommand{\L}{{\Lambda}}

\newcommand{\pr}{Phys.\ Rev.\ }

\newcommand{\prl}{Phys.\ Rev.\ Lett.\ }
\newcommand{\np}{Nucl.\ Phys.\ {\bf B}}
\newcommand{\pl}{Phys.\ Lett.\ {\bf B}}

\begin{document}
\thispagestyle{empty}
\begin{flushright}
BUHEP--94--22\\
hep-ph/9409233
\end{flushright}
\title{{\Large Physics Beyond the
Standard Model:}\\~ \\
{\Large Prospects and Perspectives}}
\vspace{2cm}
\author{R. Sekhar Chivukula
\thanks{Talk presented at DPF `94, Albuquerque, New Mexico, Aug. 2-6,
1994.}
\\
Department of Physics\\
Boston University\\
Boston, MA 02215, USA}

\maketitle

\vspace{2cm}

\begin{abstract}
        In this talk I discuss the effects of physics beyond the
standard model on the process $Z \to b\bar{b}$. I argue that,
because the top-quark is heavy, this process is susceptible to large
corrections from new physics.
\end{abstract}

\vspace{3cm}

\section{Introduction}

In terms of the painting metaphor which has been used in several of the
other theory talks at this conference, I am going to use a very narrow brush
to paint a more detailed picture of a small part of physics beyond the
standard model. In particular, instead of trying to describe all possible
constraints on all proposed models of new physics, I will concentrate on a
single process, $Z\to b\bar b$, and consider the contributions to this
process from different types of physics. I will close with some perspectives
and an advertisement for some other talks at this conference on related
topics. For a more conventional survey, I refer the reader to the
contributions of Jon Rosner \cite{rosner} and Jeff Harvey \cite{harvey}
in this conference, as well as
my talk at the Lepton-Photon conference last year \cite{sekhar}.

Let me begin by describing why I chose the process $Z\to b\bar b$. First and
foremost, it was
because of the extraordinarily precise measurement reported by the
LEP collaborations at this meeting by Richard Batley \cite{batley}:

\be
R_b = 0.2192 \pm 0.0018,
\ee

\noindent
where

\be
R_b=\frac{\Gamma(Z\rightarrow bb)}{\Gamma (Z\rightarrow hadrons)}.
\ee

\noindent
Given the reported CDF results on evidence for the top \cite{cdf}

\be
m_t = 174 \pm 10 _{-12}^{+13}
\ee

\noindent
we find a standard model prediction for $R_b$
\cite{langacker1}:

\be
R_B^{SM} = 0.2157 \pm 0.0004
\ee

\noindent
for a top in the mass range from 163 to 185 GeV. To be sure, no one would
suggest that the apparent discrepancy between the calculated value and the
reported LEP value is grounds to dismiss the standard model, especially
given that there are of the order of 25 precisely measured electroweak
quantities and only a few disagree by more than one sigma.

Nonetheless, as I will show in the rest of this talk, the $Z\to b\bar{b}$
branching ratio is particularly susceptible to contributions from new
physics. For this reason, this discrepancy though not decisive, is certainly
intriguing\footnote{For a model-independent analysis of $R_b$, see ref. 7.}.
In addition, unlike flavor-changing neutral-currents,
this process does not require GIM violation and, since it refers
to an inclusive rate, does not suffer from uncertainties due to hadronic
matrix elements or fragmentation.

\section{Standard Model}

\begin{figure}[thb]
 \vspace{1cm}
  \epsfxsize 5cm \centerline{\epsffile{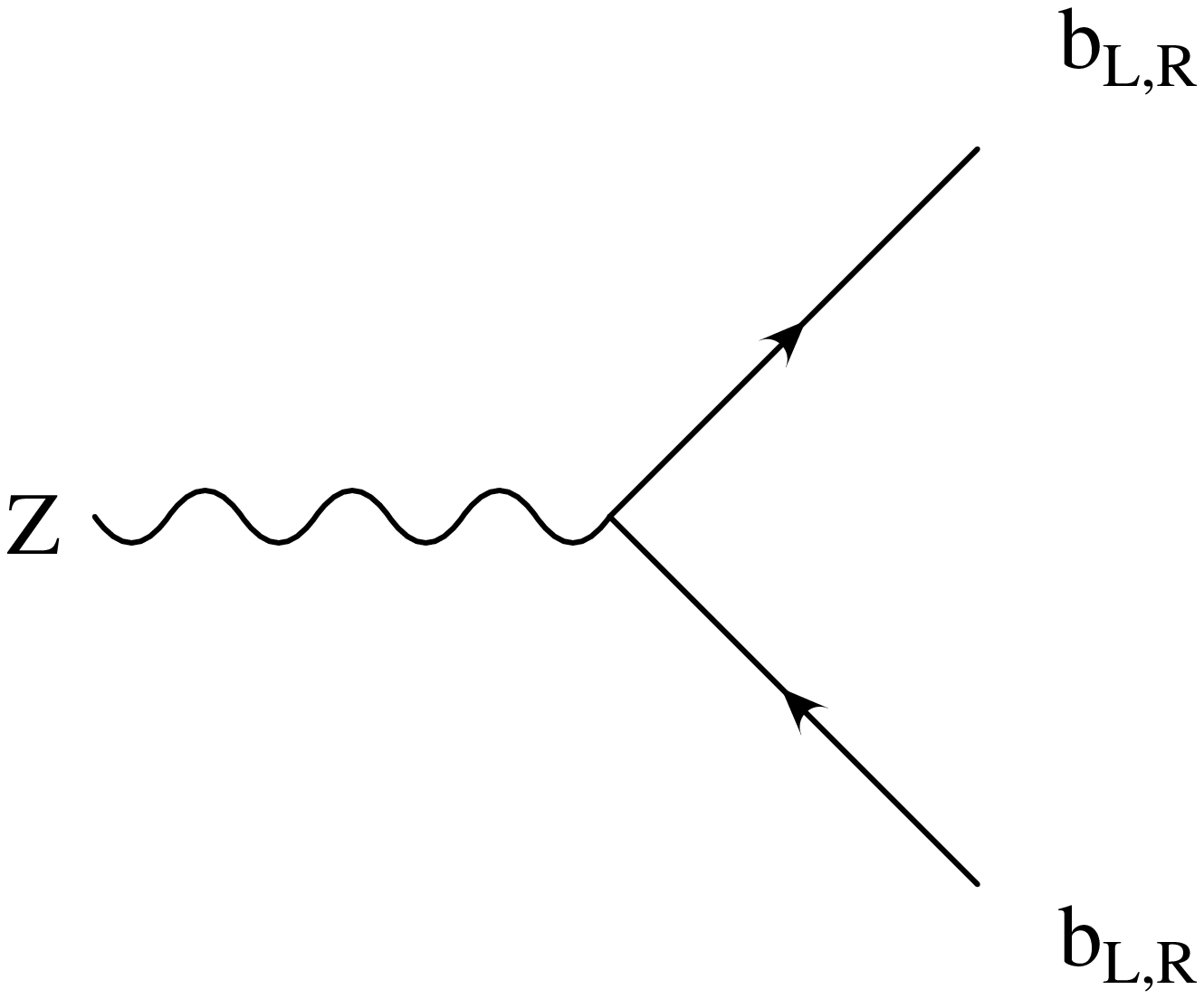}}
   \caption{~}
 \label{fig1}
\vspace{1cm}
\end{figure}

First, let us consider the process in the standard model.
At tree-level, we have the diagram shown in Fig. \ref{fig1}. The
couplings of the $b$ quark are

\be
g_L=-{\frac{1}{2}}+{\frac{1}{3}}\sin^2\theta _W
\ee

\noindent
and

\be
g_R = {\frac{1}{3}}\sin^2\theta_W.
\ee

\noindent
Using these (and the corresponding expressions for the other quarks) we find

\be
R_b^0\simeq 0.2197
\ee

\noindent
at tree-level (with $R_b$ defined as before).

\begin{figure}[htb]
\begin{center}
\begin{tabular}[t]{ccc}
\epsfxsize 3cm \epsffile{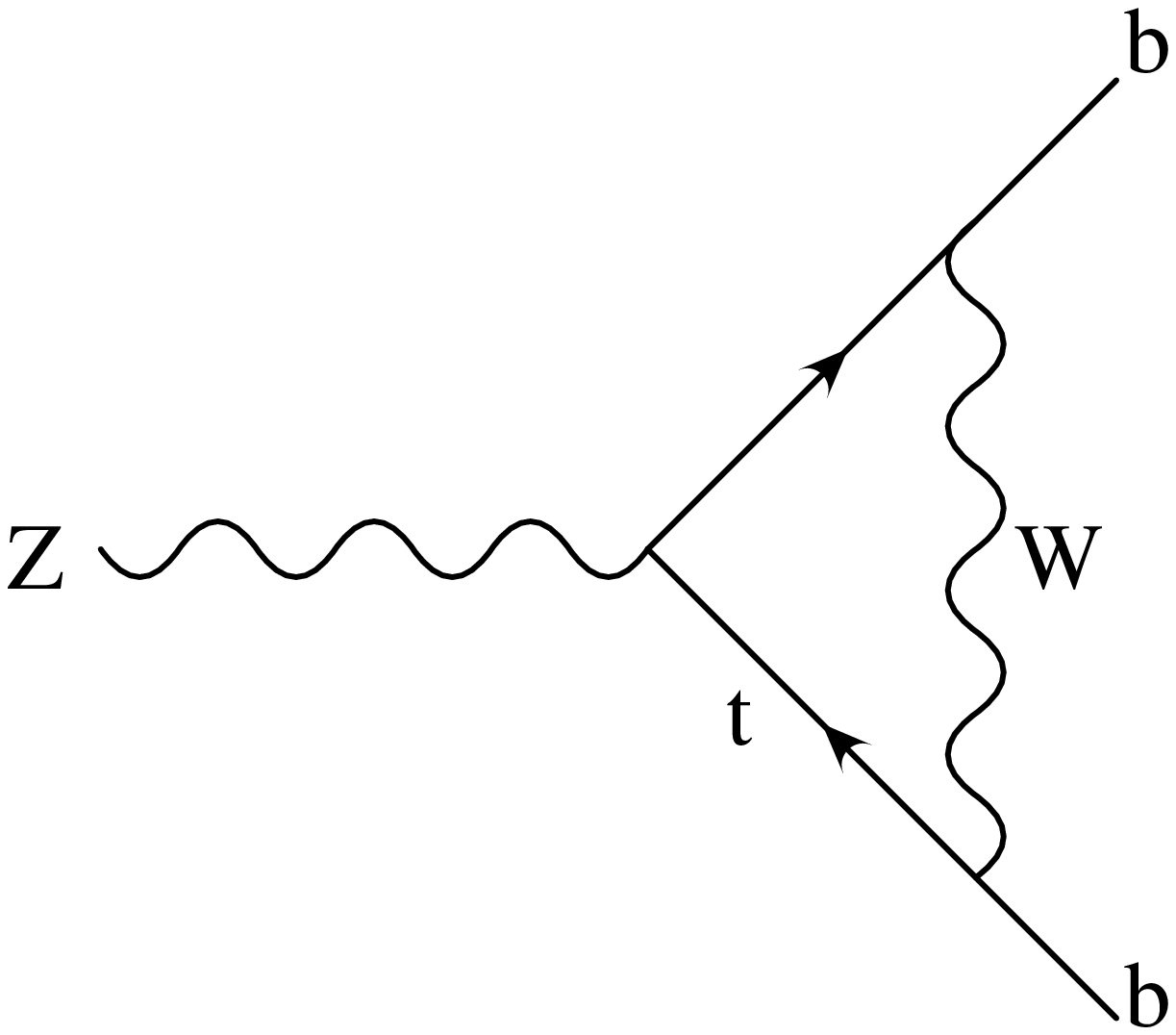}&\epsfxsize 3cm
\epsffile{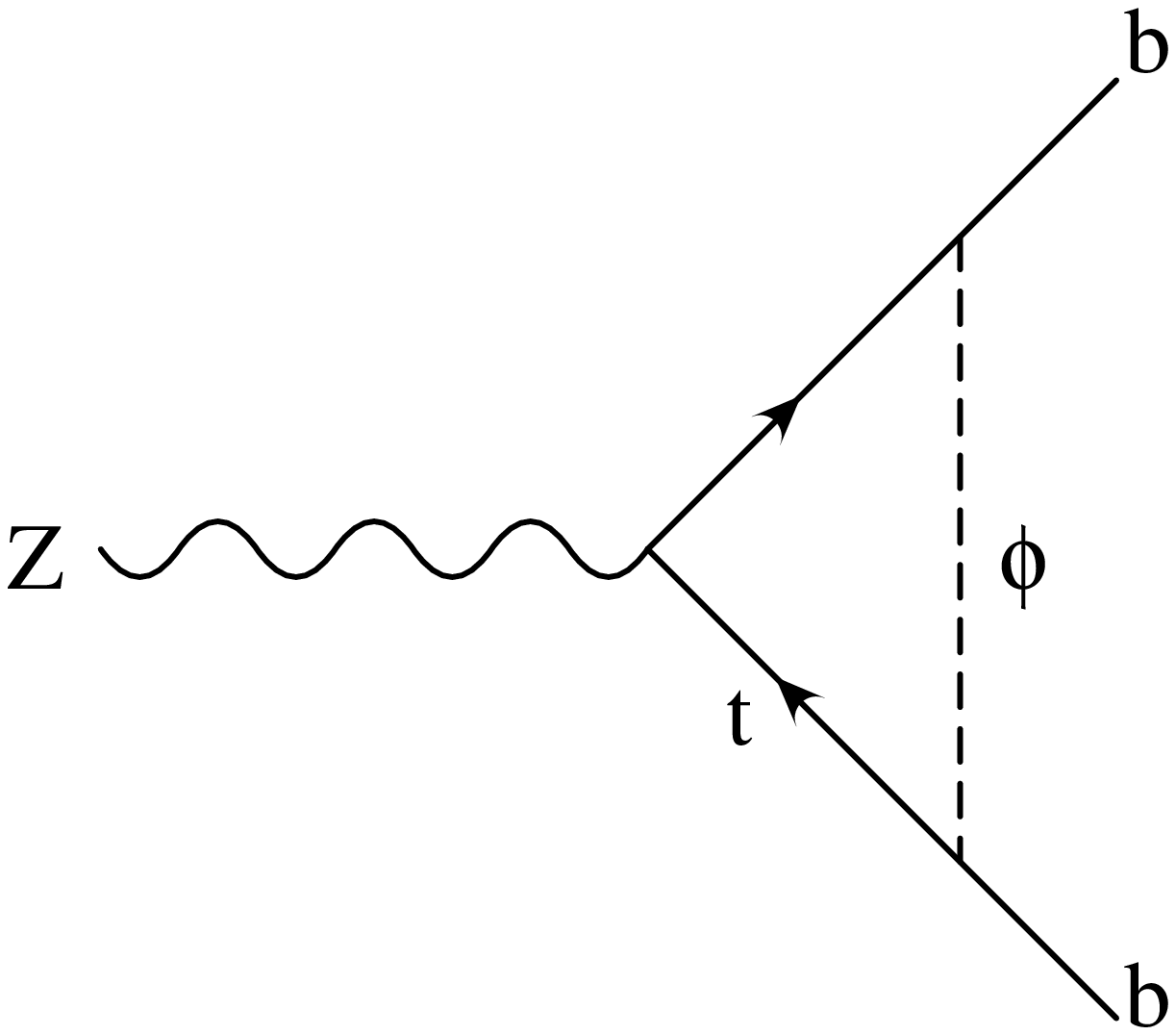}&\epsfxsize 3cm \epsffile{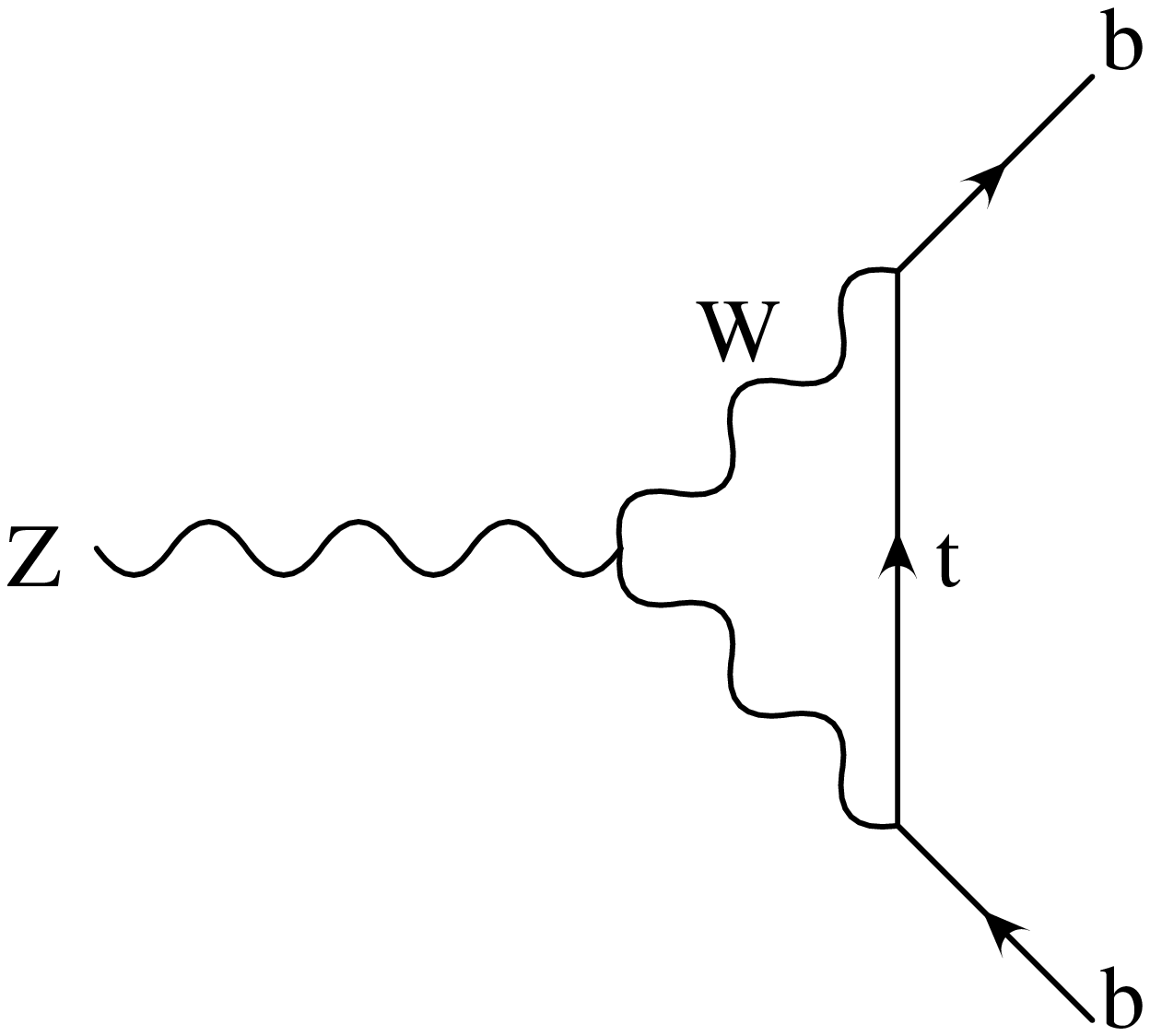}\\
\epsfxsize 3cm \epsffile{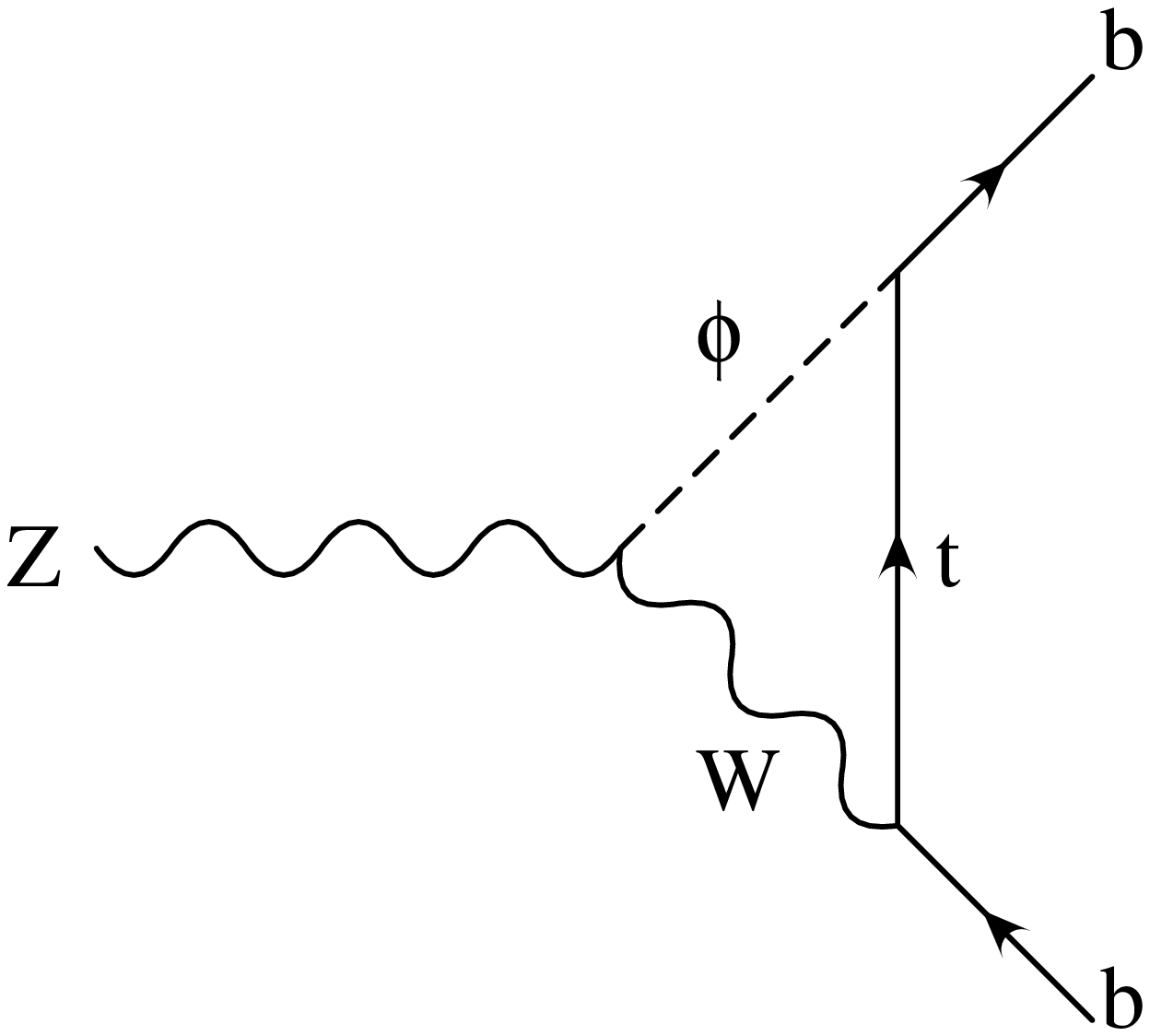}&
\epsfxsize 3cm \epsffile{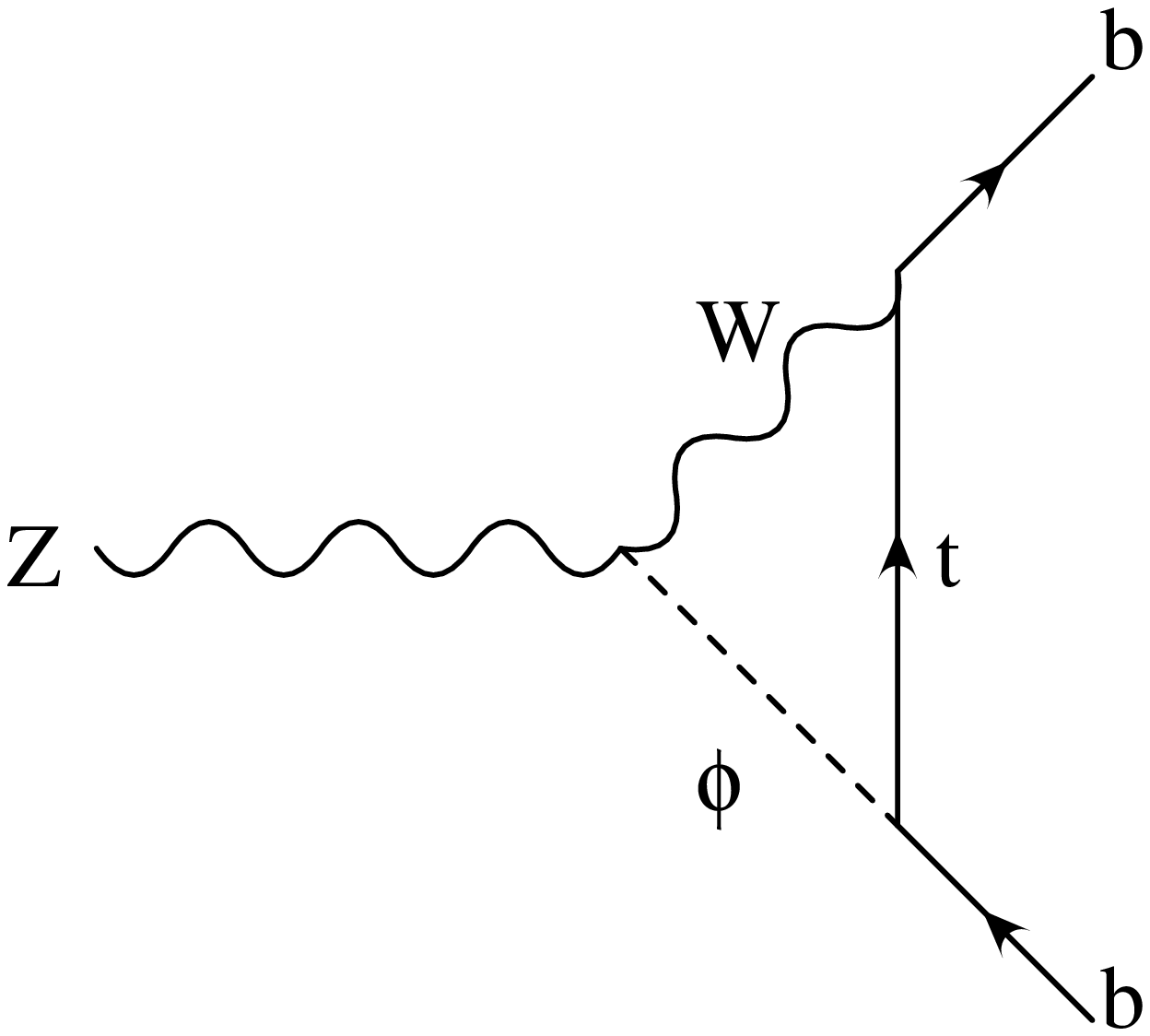}&\epsfxsize 3cm
\epsffile{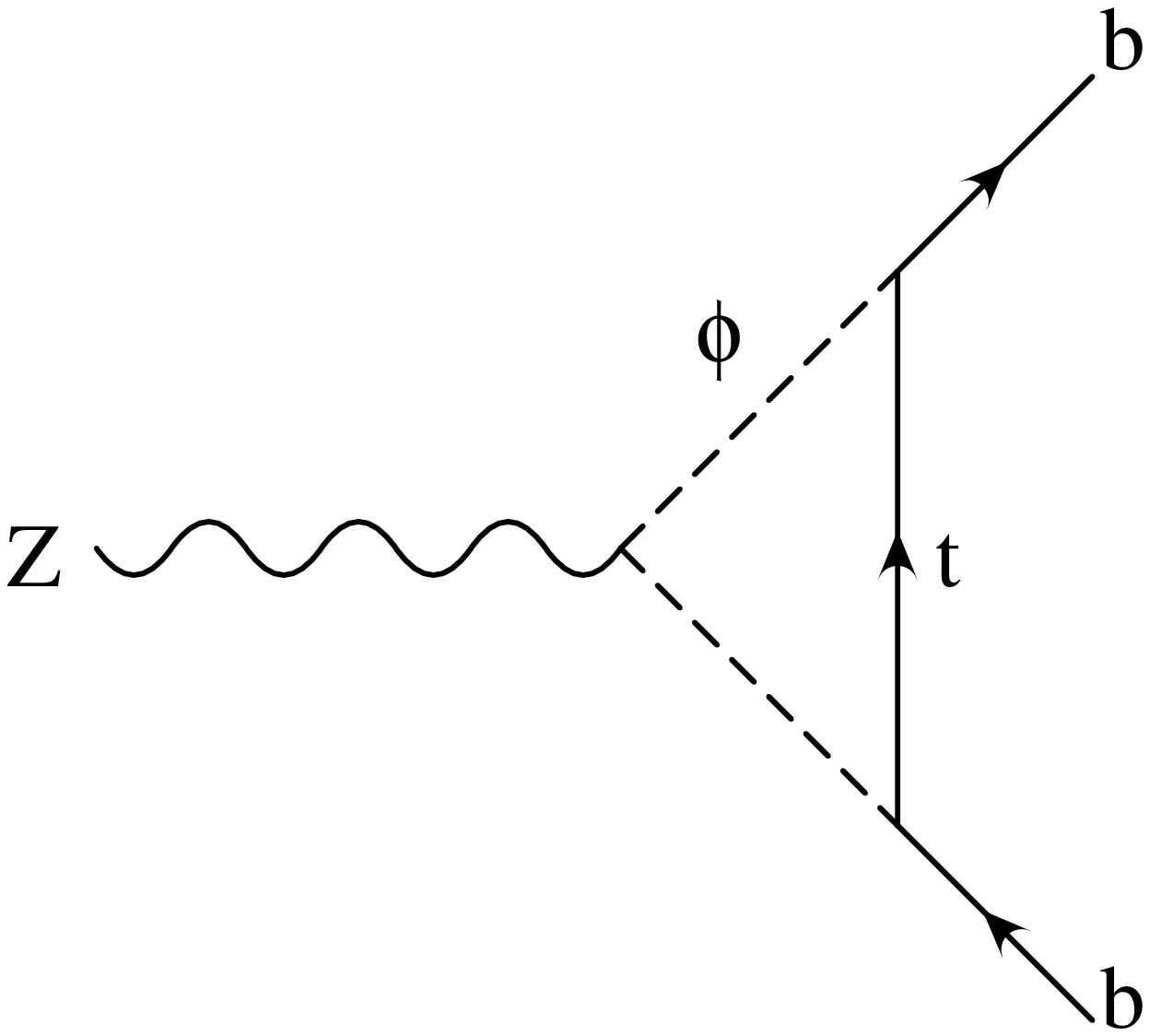}
\end{tabular}
\begin{tabular}[t]{cc}
\epsfxsize 3cm \epsffile{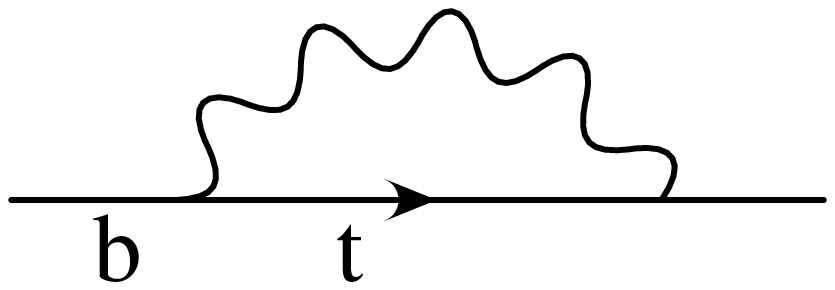}&\epsfxsize 3cm
\epsffile{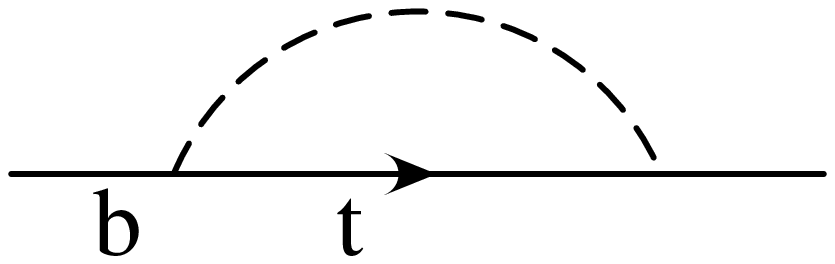}
\end{tabular}
\end{center}
\caption{ The leading standard-model corrections to $R_b$, in
t'Hooft-Feynman gauge.}
\label{fig2}
\end{figure}

\begin{figure}[htb]
\vspace{0.5cm}
\epsfxsize 12cm \centerline{\epsffile{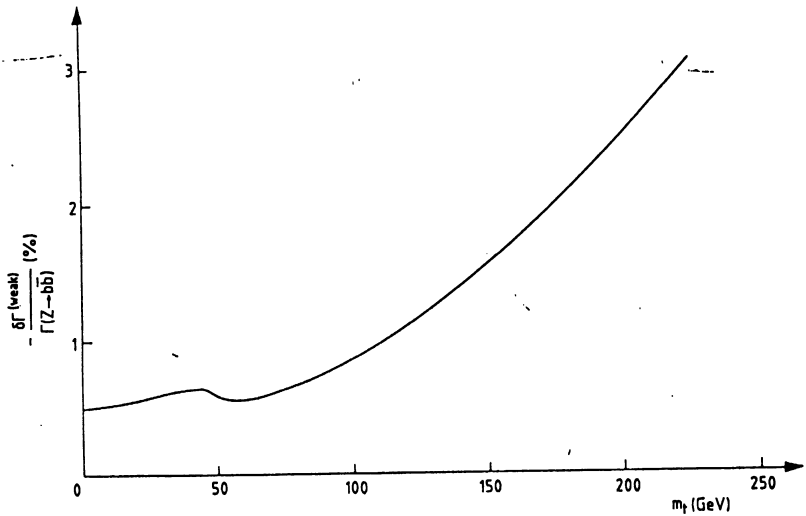}}
\caption{ Fractional change in $\Gamma(Z \to b\bar{b}$ (in \%)
as a function of $m_t$ (in GeV). From ref. 11.}
\label{diez}
\vspace{0.5cm}
\end{figure}

As with other precisely-measured quantities at LEP, we also need to consider
the leading (one-loop) radiative corrections to this quantity. The
advantages of the {\it ratio} $R_b$ now become clear: both the
flavor-independent (``oblique'') corrections \cite{lynn} and
the leading
QCD corrections (which together are generally the most important radiative
corrections) largely cancel in this
ratio \cite{benneloch}.
Therefore, the leading corrections to $R_b$ are the {\it non-universal}
corrections to the $Z\to b\bar b$ vertex. In t'Hooft--Feynman gauge,
these vertex corrections, along
with the corresponding wave-function renormalization diagrams, are shown in
Fig. \ref{fig2} \cite{riemann,bernabeu}.
The results of this computation, shown as a fractional change in the partial
width of the $Z$ to $b$ quarks, is shown in Fig. \ref{diez}
\cite{bernabeu}. As we see, this
correction varies from a little less than 1.5\% to a little less than 2.5\%
as the top mass varies from 150 to 200 GeV.

In the limit $m_b\to 0$, there is no change in the right-handed $b$ quark
coupling, and the result of the calculation may be written

\be
\delta g_L^b=A{\frac{m_t^2}{16\pi ^2v^2}}+B
{\frac{g^2}{16\pi ^2}}\log ({\frac{m_t}{M_W}})^2+\label{pedro}
\cdots\footnote{Note that while such an expansion in (inverse) powers of the
top
quark mass is useful for the purposes of illustration, one must go
to quite high order in order to obtain an accurate result \cite{bernabeu}.}
\ee

\noindent
where $A$ and $B$ are computable constants.

\begin{figure}[htb]
\vspace{1cm}
\epsfxsize 4cm \centerline{\epsffile{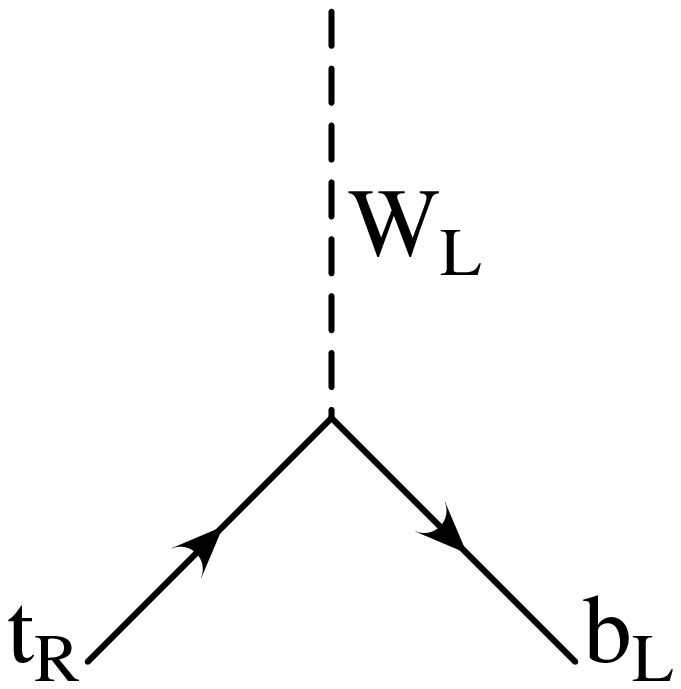}}
\caption{~}
\label{tres}
\vspace{1cm}
\end{figure}

Some features of the standard model calculation are of particular note.
First, the result
does not go to zero as $m_t\to \infty $. That is,
the contribution does not decouple in $m_t$. The reason for this is that the
couplings of the unphysical Goldstone bosons (and more generally of
the longitudinal gauge bosons)
to the
$t_R$ and $b_L$, Fig. \ref{tres}, are proportional to $m_t$. Hence the
Appelquist-Carazzone decoupling theorem \cite{carazzone} does not apply.


The fact that this coupling is proportional to $m_t$ is not restricted to
the standard model. As emphasized by Peccei and Zhang\cite{zhang},
this result follows from the electroweak generalization of the
Goldberger-Treiman (GT) relation \cite{gt}. In QCD, the GT relation reads

\be
g_{\pi NN}=\frac{g_Am_N}{f_\pi }
\label{gt}
\ee

\noindent
where $g_{\pi NN}$ is the pion-nucleon coupling, $m_N$ the mass of the
nucleon, $f_\pi $ the pion decay constant, and $g_A$ is the renormalization
of the axial-vector couplings (approximately 1.25 in QCD). In the
electroweak theory, this relation reads

\be
g_{W_Lt_Rb_L}=\sqrt{2}\frac{g_Am_t}{v},
\ee

\noindent
where $v\approx 246$ GeV, and $g_A=1$ at tree-level in
the standard model. Non-decoupling contributions appear in {\it all}
theories.

\begin{figure}[htb]
\begin{center}
\begin{tabular}{ccc}
\epsfxsize 4cm \epsffile{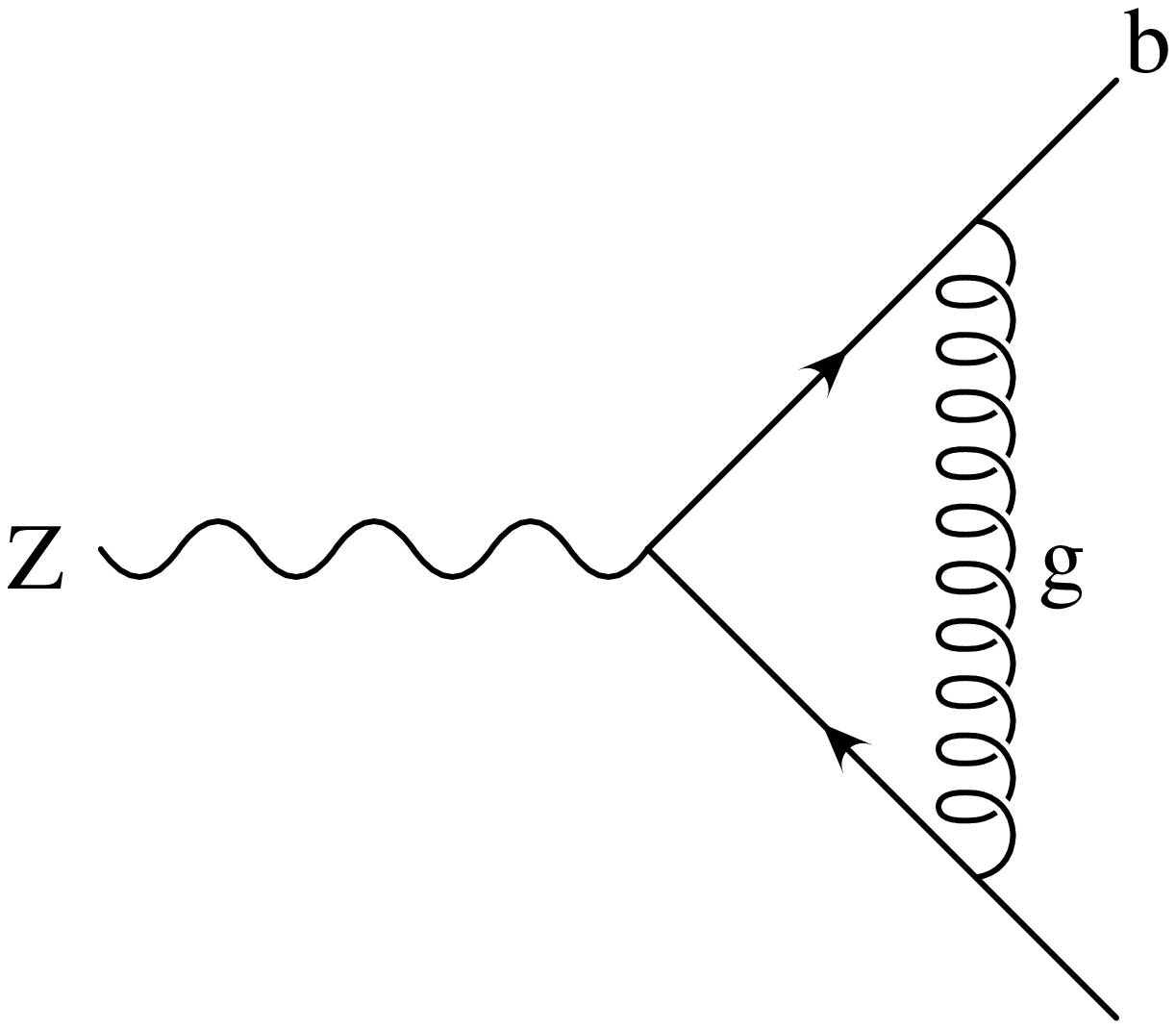}&\hspace{1cm}
\epsfxsize 4cm \epsffile{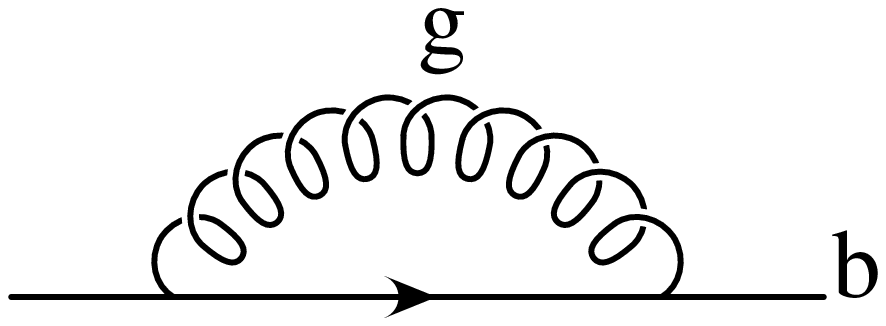}
\end{tabular}
\end{center}
\caption{Potential strong corrections to $R_b$ which vanish in the
limit $m_b \to 0$.}
\label{cuatro}
\end{figure}

Finally, we note that in order to have $\delta g_L\neq 0$ we must have
$SU(2)\times U(1)$ breaking. In an unbroken gauge theory, the gauge currents
are not renormalized: this, after all, is the reason why the $\bar p$ and $e$
charges are the same -- independent of the effects of QCD. In the absence
of electroweak symmetry breaking, the current to which the $Z$ couples {\it %
cannot} be renormalized and $R_b$ would
not change. This is why (in the limit $m_b\to 0$) there are no strong
vertex corrections like the ones depicted
in Fig. \ref{cuatro}

\section{Two Scalar Doublet Models}

The simplest extension to the standard model is one in which the electroweak
symmetry breaking sector involves two fundamental scalar doublets, $\varphi
_1$ and $\varphi _2$, instead of one \footnote{For
a general review of these models, see ref. 15.}.
The new scalar degrees of freedom result in the appearance of an extra
pair of charged scalar particles, $H^{\pm }$, as well as a pseudo-scalar and
an additional scalar particle. The expectation values of the two scalars may
be written

\be
\langle \varphi _1\rangle=\left(
\begin{array}{c}
v_1\\ 0 \\
\end{array}\right)
\ee

\noindent
and

\be
\langle \varphi _2\rangle=\left(
\begin{array}{c}
v_2\\ 0 \\
\end{array}\right)
\ee

\noindent
In order for the $W$ and $Z$ masses to be correct, we require that

\be
v_1^2+v_2^2=v^2
\ee

\noindent
where $v$ is the usual weak scale. Given the relation above, it is natural
to define an angle $\beta $ such that

\be
v_1=v\cos \beta \ \ \
v_2=v\sin \beta .
\ee

\noindent
Then, the relationship between the charged scalar fields in the mass
eigenstate fields is

\be
\pi ^{\pm }=\cos \beta~
\varphi _1^{\pm }+\sin \beta~ \varphi _2^{\pm }
\ee

\noindent
for the ``eaten'' Goldstone boson and

\be
H^{\pm }=-\sin \beta~
\varphi _1^{\pm }+\cos \beta~ \varphi _2^{\pm }
\ee

\noindent
for the extra physical charged scalars.

Conventionally, it is expected that only one of the original scalar doublets
(which we take to be $\phi_1$)
couples to the $t_R$ so as to avoid flavor-changing neutral-currents
\footnote{Though
this may not be strictly necessary \cite{hall}.}. This results in the
couplings

\be
{\frac{m_t}{v_2}}\bar
t_R\varphi_1^{+}b_L\to {\frac{m_t}{v\sin \beta }}\bar t_R[
\pi^{+}\sin \beta +H^{+}\cos \beta ]b_L
\ee

\noindent
to the mass eigenstate fields. Examining this expression, we see that the
Goldstone boson field $\pi ^{+}$ couples to $\bar t_Rb_L$ with the same
strength as the standard model, while the coupling of the $H^{+}$ differs
from this by a factor of $\cot \beta $.
Since the coupling of the Goldstone boson field is the same as in the
standard model, the calculations of the previous section still apply. This
is a general result: unlike many weak radiative corrections, in the limit
$m_b\to 0$ the standard
model correction to the $Z b\bar{b}$ vertex does not involve the Higgs
boson, only the longitudinal gauge bosons. And therefore, to the extent that
$g_A \approx 1$, these contributions arise in {\em all} theories.

\begin{figure}[htb]
\begin{center}
\begin{tabular}{ccc}
\epsfxsize 4cm \epsffile{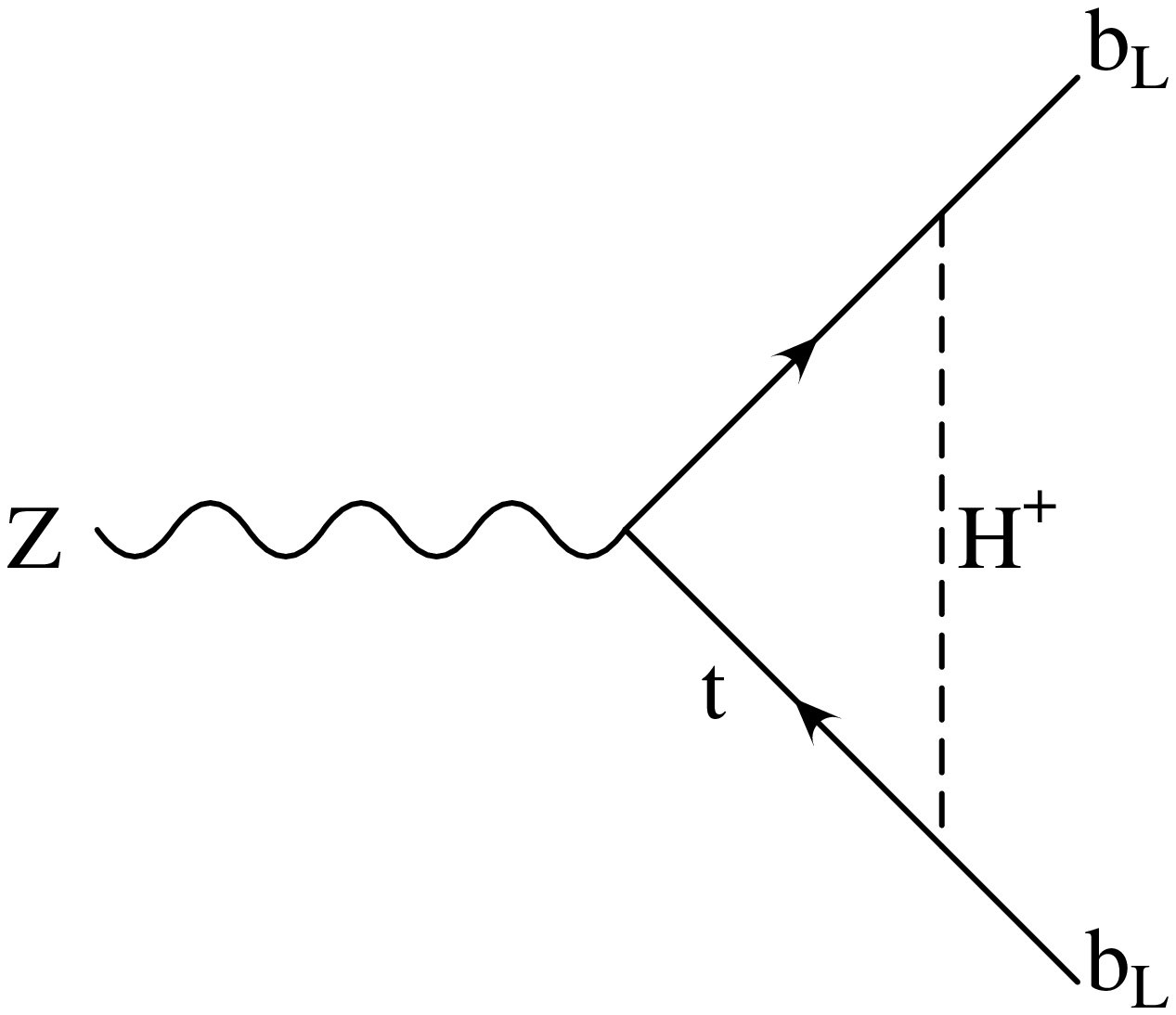}&
\epsfxsize 4cm \epsffile{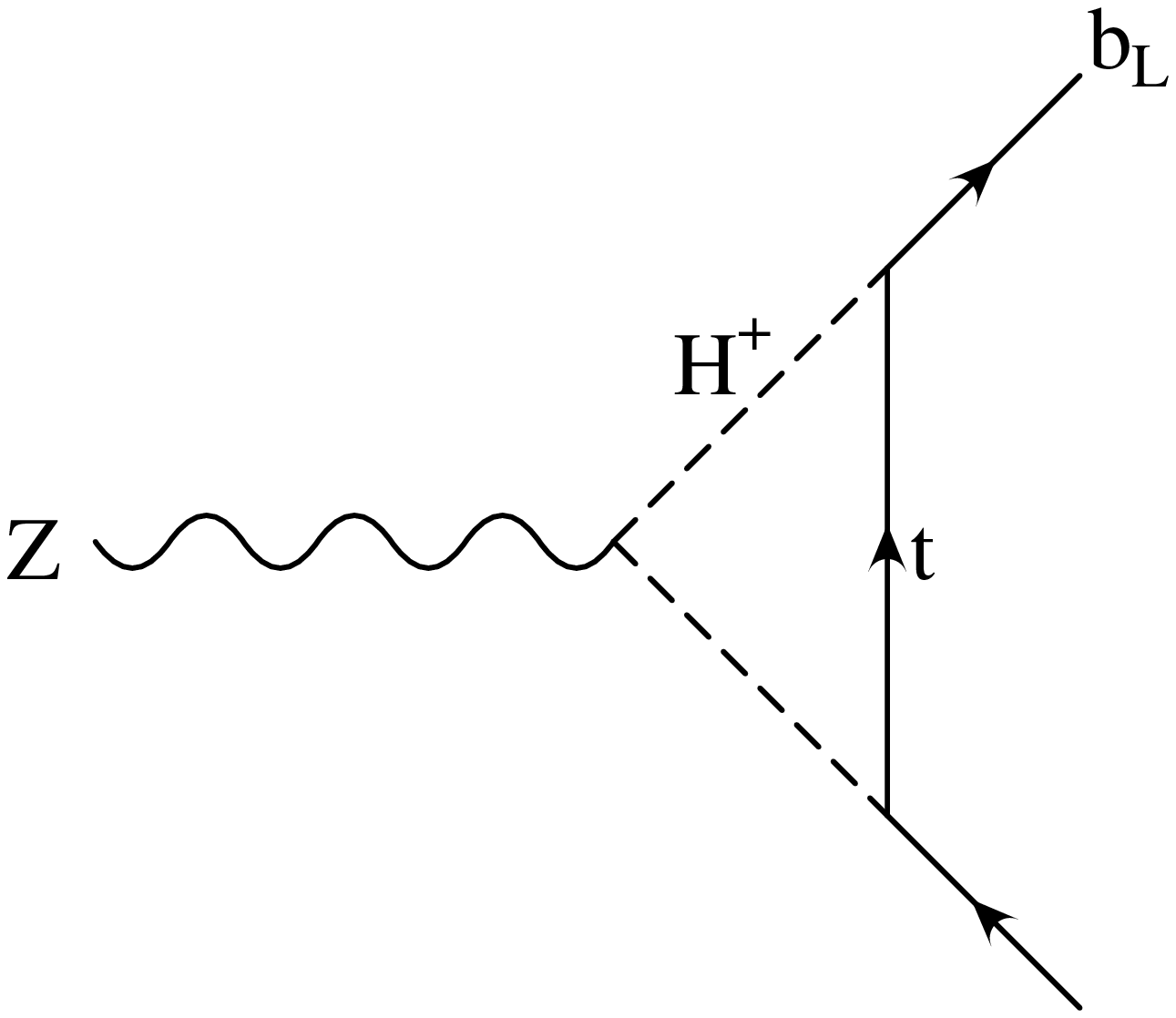}&
\epsfxsize 3cm \epsffile{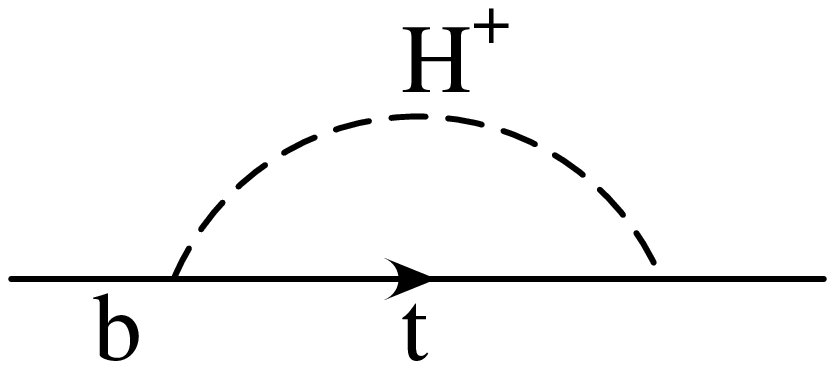}
\end{tabular}
\end{center}
\caption{Corrections to $R_b$ in the two-scalar doublet models.}
\label{seis}
\end{figure}

There are, however, additional contributions coming from the exchange of the
extra charged scalars. These corrections are
shown in Fig. \ref{seis}.

\begin{figure}[htb]
\vspace{0.5cm}
\epsfxsize 12cm \centerline{\epsffile{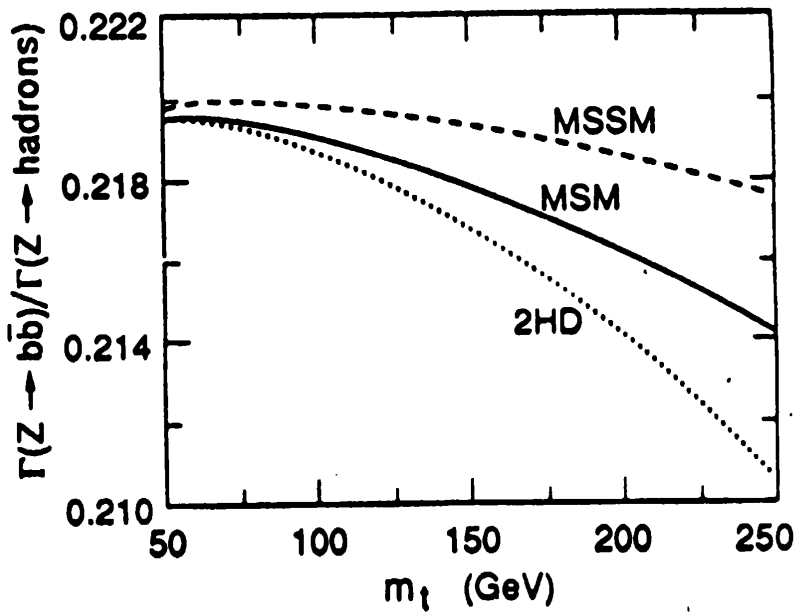}}
\caption{$R_b$ as a function of $m_t$ (in GeV) in the standard
model (MSM), two-scalar doublet model (2HD), and the minimal
supersymmetric standard model (MSSM), assuming $\tan \beta = 1$,
$m_{\widetilde{t}}=M_{H^+}=100$ GeV, $\mu=30$ GeV
and $M=50$ GeV. From ref. 18.}
\label{graph2}\vspace{0.5cm}
\end{figure}

\noindent
Note that these diagrams are a subset of the diagrams shown in Fig. \ref{fig2},
with
the replacement $\pi ^{+}\to H^{+}$, resulting in the couplings changing by
a factor of $\cot {}^2\beta $ and the replacement of $M_W$ by $M_{H^{+}}$.
For $\tan \beta \approx 1$ and $M_{H^{+}}\approx M_W$ therefore, we expect
an effect of the same order of magnitude as the standard model \cite{hollik}.
This
effect is shown in Fig. \ref{graph2}\cite{tota}. \footnote{These values of
$\tan\beta$ and
$M_{H^+}$ are chosen for the purposes of illustration only. Recent
results from CLEO on $b \to s \gamma$ require that the charged
scalar mass be greater than about 230 GeV \cite{cleo}.}

Note that, as in
the standard model, this effect tends to {\it reduce} the width of the $Z$
to $b\bar b$. This tendency holds in all two-scalar doublet models except
in the limit where $\tan \beta $ is {\it very} large: there the Yukawa
coupling of the $b$ quark can be comparable to that of the $t$ quark.
Processes
involving intermediate $b$ quarks and neutral scalars become important, and
can result in an {\it increase} of $R_b$ \cite{student}.

Two features of this calculation are of
particular note. First, because the Yukawa
coupling of the charged scalar is proportional to $m_t$,

\be
\lambda_t \sim \frac{m_t}
{v\tan \beta },
\ee

\noindent
the effect on $R_b$
does not decouple in $m_t$.
Second, the effect on $R_b$ {\it does} vanish in the limit that $%
m_{H^{+}}\to \infty $. This is because there is an $SU(2)\times U(1)$
preserving mass term for the two scalar doublets,

\be
-\mu ^2(\varphi _1^{\dagger
}\varphi _2+h.c.)
\ee

\noindent
which can be introduced in the Lagrangian. In the limit that $\mu ^2\to
\infty $, the theory reduces precisely to the standard model. For this
reason, the extra contributions can be made {\it arbitrarily small},
independent of the $t$ and $W$ masses.

\section{Supersymmetry}

To judge by the volume of submissions on hep-ph, the most popular extensions
of the standard model involve low-energy
supersymmetry\footnote{For a review, see ref. 20.}. In the
minimal version of this scenario, one introduces superpartners (a fermionic
partner for every boson and visa versa) for all of the ordinary standard
model particles

\be
 \begin{array}{ll}
q_L & \to
\widetilde{q}_L \\ u_R & \to
\widetilde{u}_R \\ d_R & \to
\widetilde{d}_R \\ l_L & \to
\widetilde{l}_L \\ e_R & \to
\widetilde{e}_R \\ g & \to
\widetilde{g} \\ W^{\pm} & \to
\widetilde{W}^{\pm} \\ Z & \to
\widetilde{Z} \\ {\gamma} & \to
\widetilde{\g}
\end{array}
\ee

\noindent
In addition supersymmetry requires that the theory involve (at least) two
weak-doublet chiral superfields to perform the role of the standard model
Higgs doublet.

\be
\begin{array}{ll}
H_1 & \to
\widetilde{H}_1 \\ H_2 & \to
\widetilde{H}_2 \\  &
\end{array}
\ee

\begin{figure}[htb]
\begin{center}
\begin{tabular}{cl}
\epsfxsize 8cm \epsffile{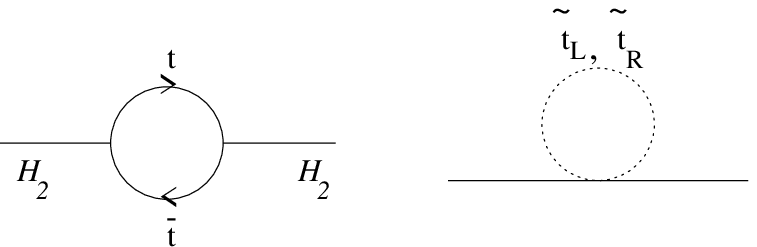}&
$\begin{array}{l}
\sim \frac{-3}{16\pi^2}\l_t^2 M^2_{
\widetilde{Q}} \log\left(\frac{\L^2}
{M^2_{\widetilde{Q}}}\right)\\ ~\vspace{1.5cm}~
\end{array}$.
\end{tabular}
\end{center}
\caption{~}
\label{pepa}
\end{figure}

The primary attraction of supersymmetric theories is that corrections to the
Higgs mass are no longer quadratically dependent on the cutoff, as we
see (for example) in Fig. \ref{pepa}.

\noindent
Quadratic divergences are absent because the mass of the Higgs boson is related
by supersymmetry
to the mass of its fermionic partner, and the mass of this fermionic partner
can be protected by a chiral symmetry. A light Higgs can be (technically)
natural in SUSY.

Of course, SUSY cannot be exact. None of the extra particles required by
supersymmetry have been observed. If SUSY is broken softly, the symmetry
breaking does not reintroduce the quadratic divergences of an ordinary
fundamental scalar theory. In a theory with soft SUSY breaking, the
radiative corrections to the Higgs masses end up being proportional to the
masses of the SUSY partners. Since we want the Higgs to ``naturally'' have a
mass of order 1 TeV, SUSY is relevant to the
hierarchy problem if the masses of the superpartners are of order 1
TeV (or less).

\begin{figure}[htb]
\begin{center}
\begin{tabular}{ccc}
\epsfxsize 4cm \epsffile{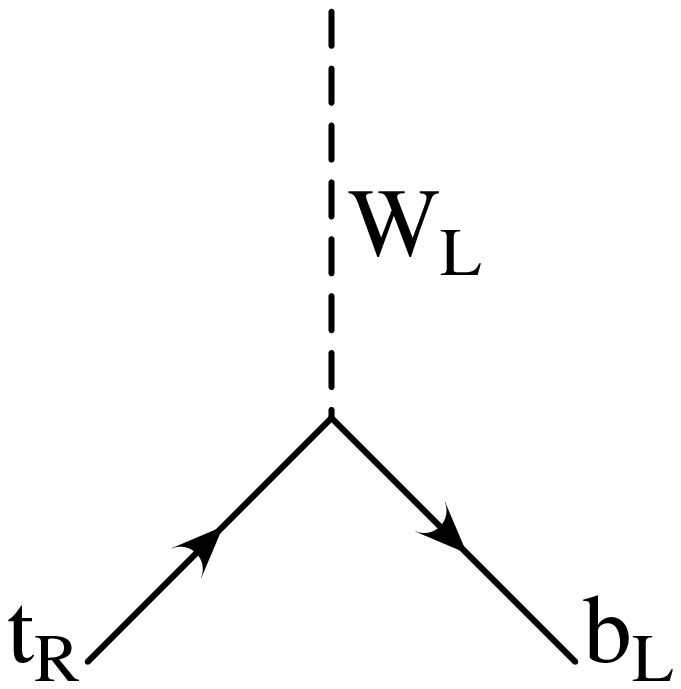}&
\epsfxsize 2cm \epsfbox[212.4 311.76 418.975 513]{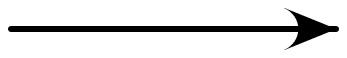}&
\epsfxsize 4cm \epsffile{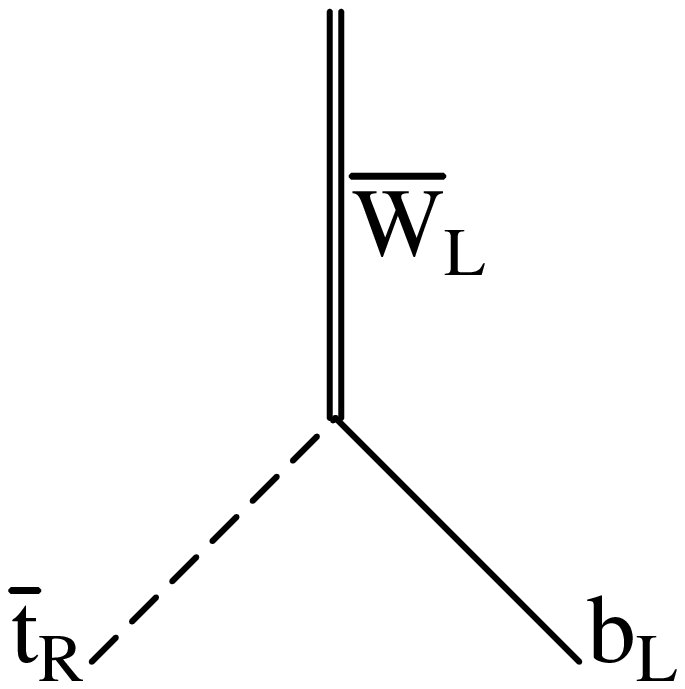}\\
\epsfxsize 4cm \epsffile{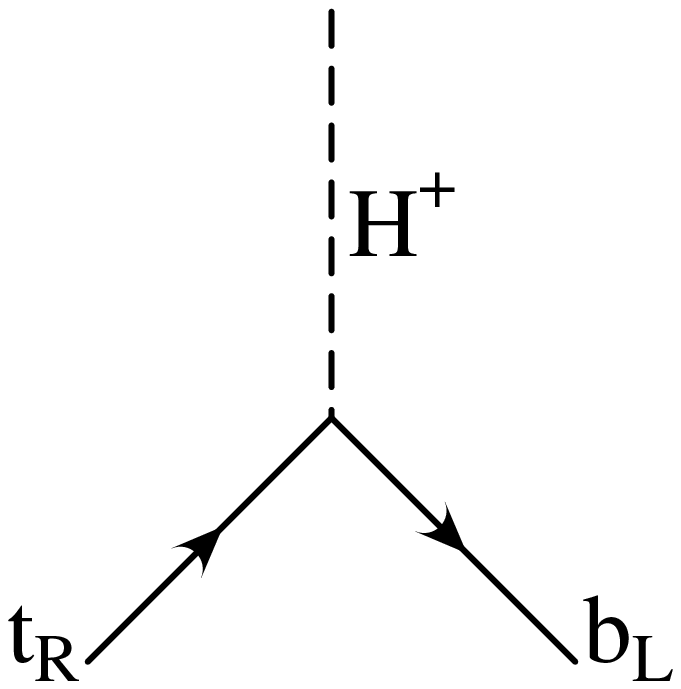}&
\epsfxsize 2cm \epsfbox[212.4 311.76 418.975 513]{fig.ps}&
\epsfxsize 4cm \epsffile{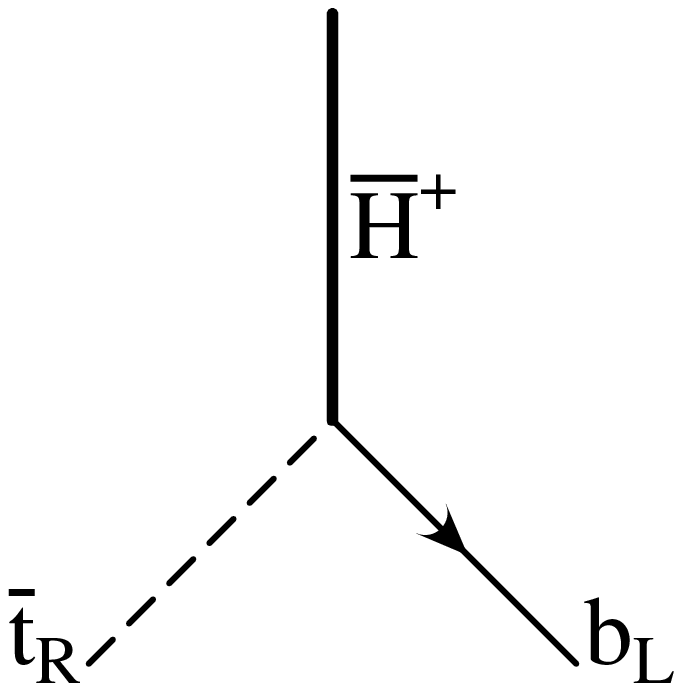}
\end{tabular}
\end{center}
\caption{ SUSY interactions which will contribute to $R_b$.}
\label{jose}
\end{figure}

\begin{figure}[htb]
\begin{center}
\begin{tabular}{ccc}
\epsfxsize 4cm \epsffile{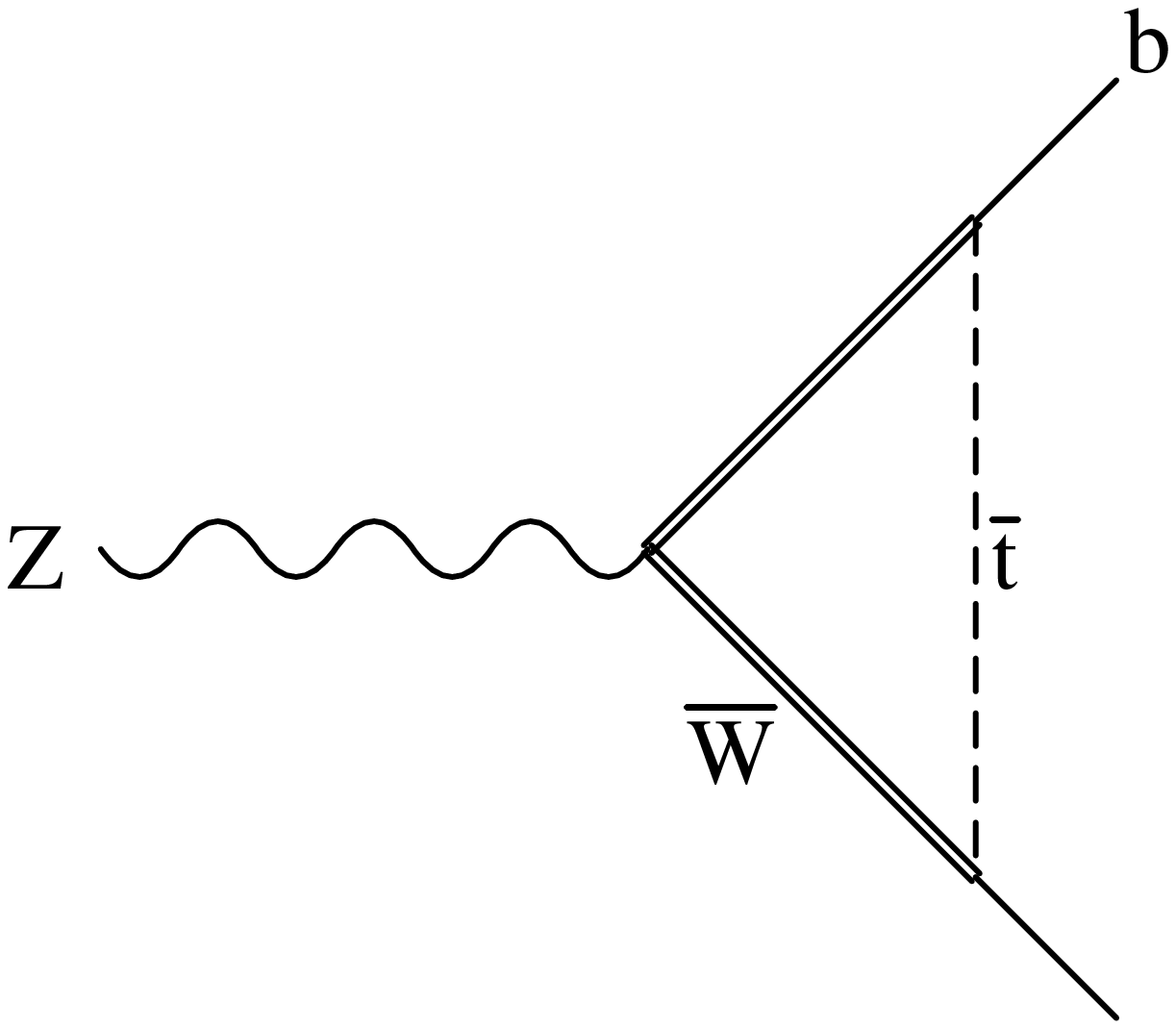}&
\epsfxsize 4cm \epsffile{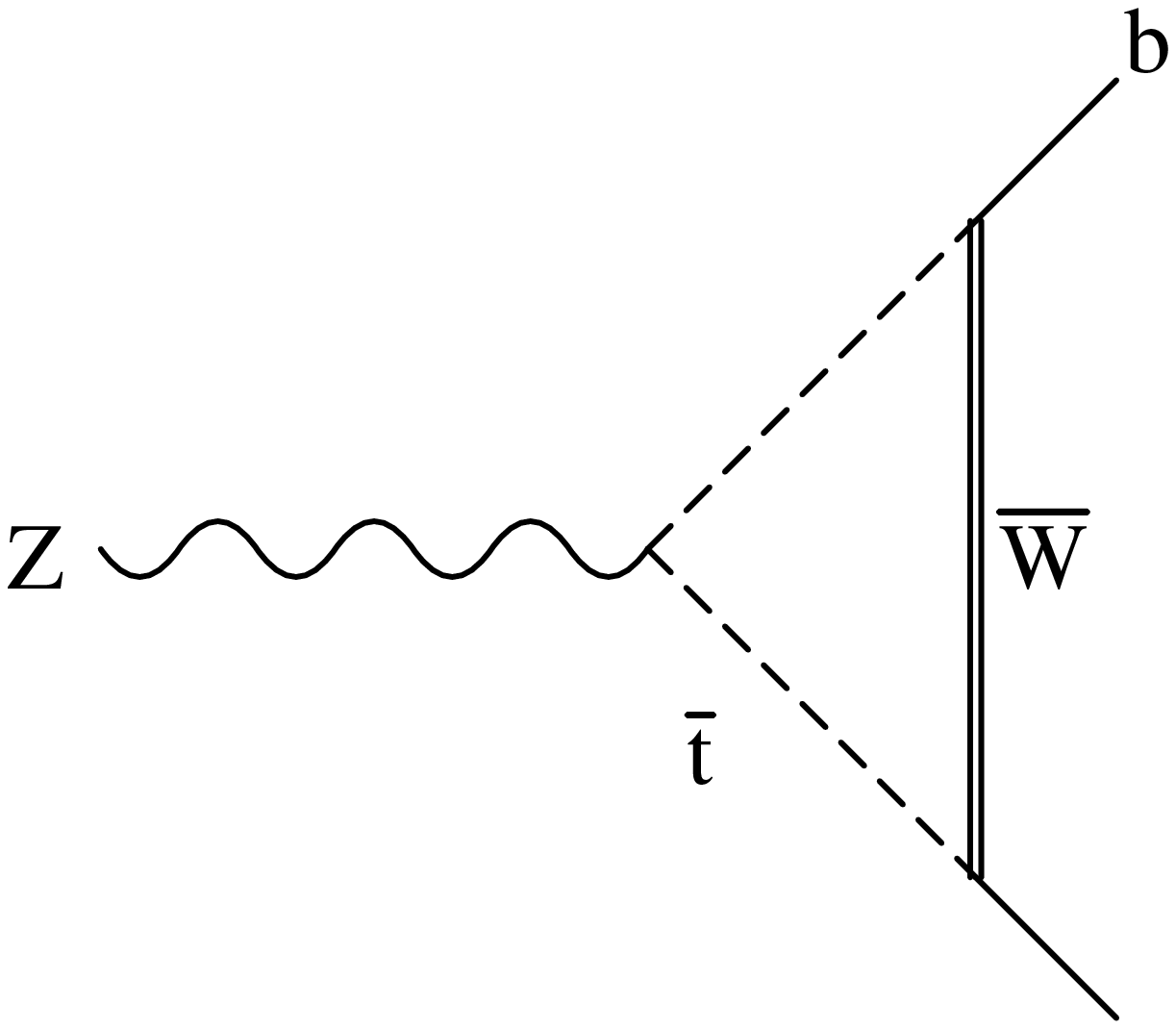}&
\epsfxsize 4cm \epsffile{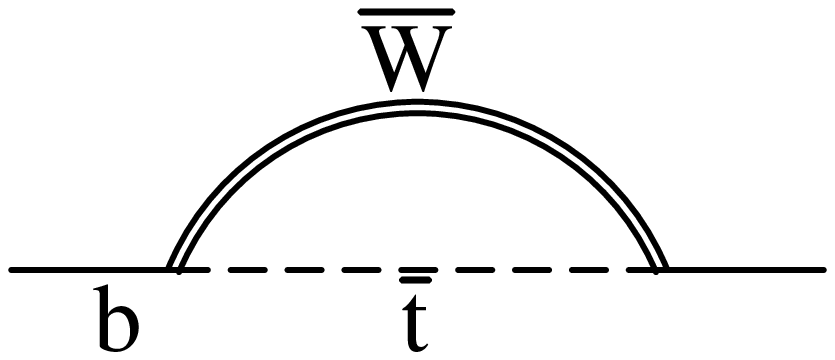}\\
\epsfxsize 4cm \epsffile{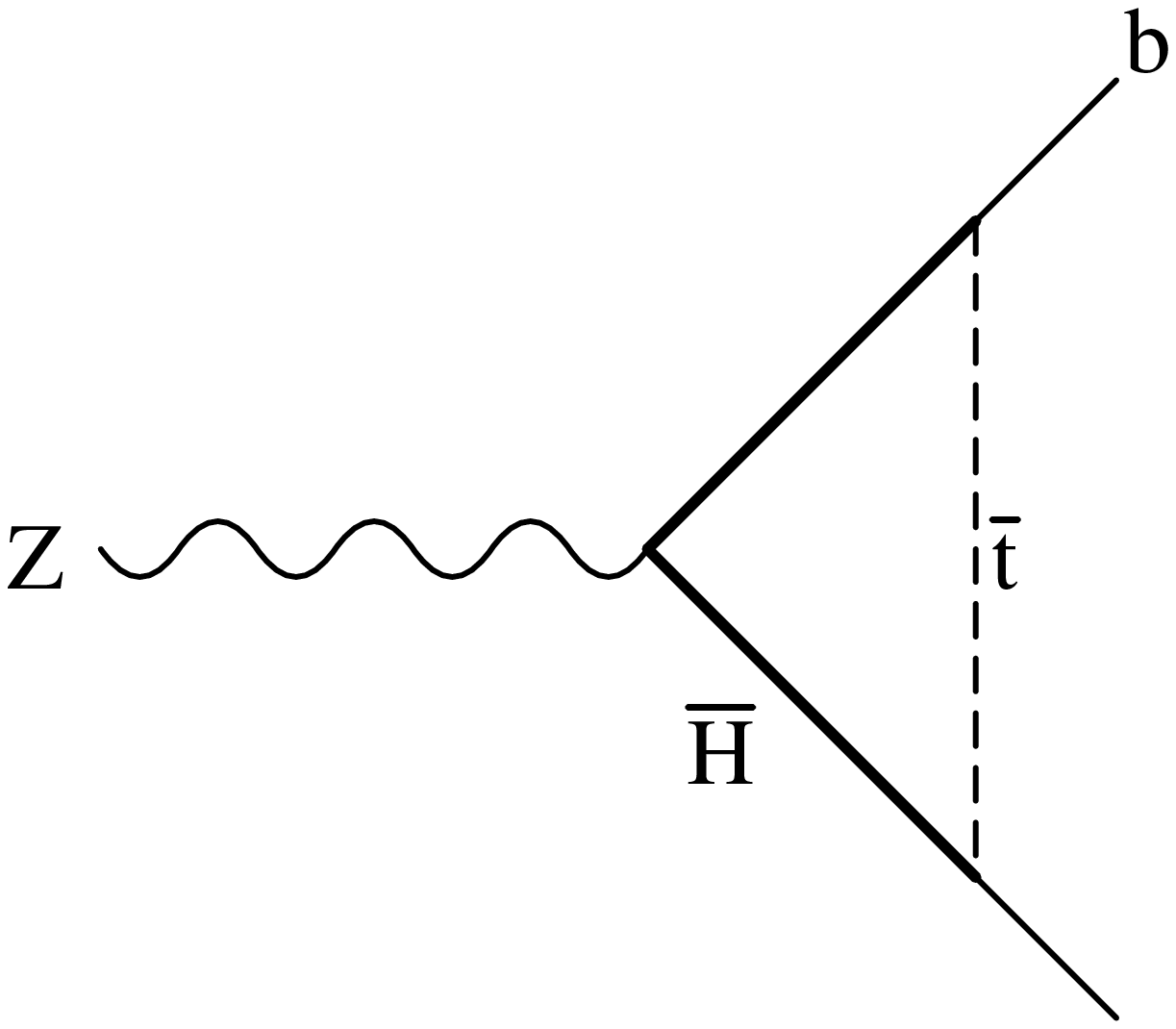}&
\epsfxsize 4cm \epsffile{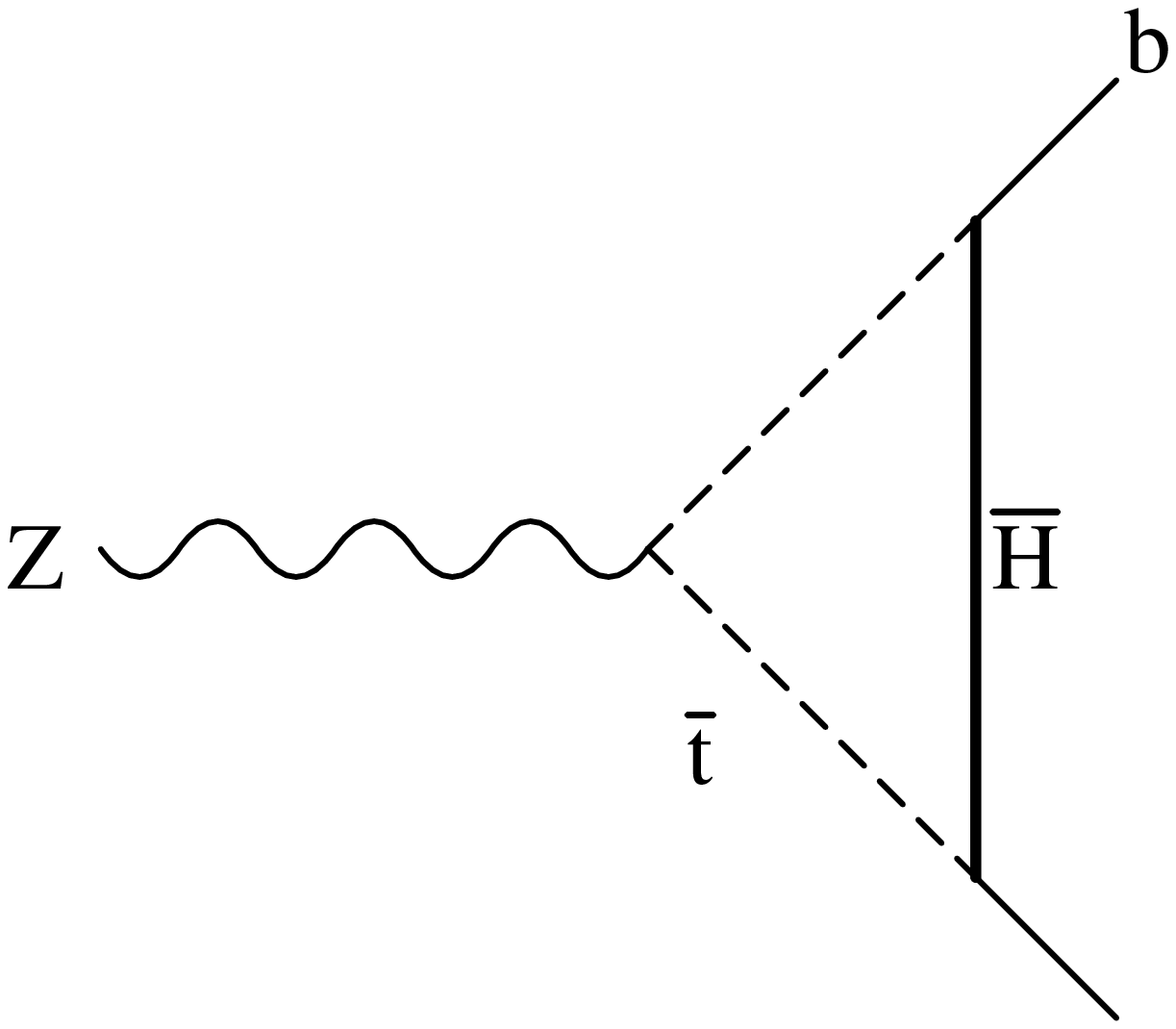}&
\epsfxsize 4cm \epsffile{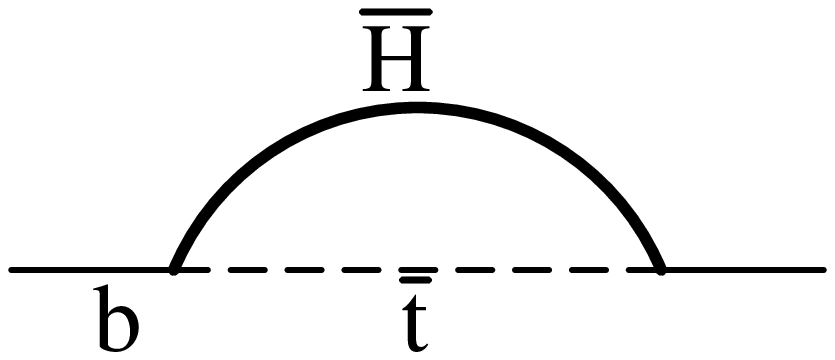}
\end{tabular}
\end{center}
\caption{ Corrections to $R_b$ in supersymmetric models.}
\label{josefa}
\end{figure}

In SUSY theories, in addition to the contributions discussed in the last two
sections, we have contributions coming from intermediate states involving
the superpartners. The relevant vertices are
shown in Fig. \ref{jose} and the new
contributions in Fig. \ref{josefa}. Notice that
the first set of vertices in Fig.
\ref{jose} are proportional to $m_t/v$ while the second are proportional
to $m_t/v\tan\beta$.
For a particular choice of superpartner and Higgs masses,
the results of this computation are plotted in
Fig. \ref{graph2}. As
shown, for those relatively light superpartner masses (of order $M_W$)
the result is of
the same order of magnitude as the correction in the standard model, but has
the {\it opposite} sign: the effects of radiative corrections involving
superpartners tend to {\it increase} $R_b$.

In terms of the analysis presented before, these couplings are
non-decoupling in $m_t$, but decoupling in the superpartner (top squark \&
chargino) masses. In the limit where the superpartner masses are large, but
the charged-scalar masses are small, the total effect on $R_b$ can approach
that of the two-scalar model presented in the last section. The overall
contribution, therefore, could be anywhere between the two-scalar and MSSM
contributions shown on Fig. \ref{graph2} \cite{tota}.

\begin{figure}[htb]
\vspace{0.5cm}
\epsfxsize 10cm \centerline{\epsffile{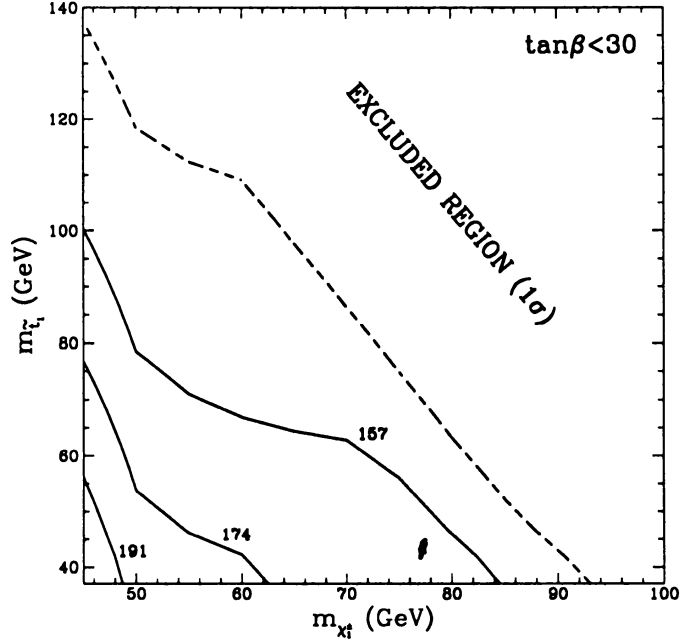}}
\caption{One-sigma limits on chargino and top-squark masses coming from the
measured value of $R_b$ for various (191, 174, \& 157 GeV) top-quark
masses. The dashed line represents the upper-bound for a top-quark
mass of 174 GeV and $R_b \ge 0.2172$. From ref. 21.}
\label{graph1}\vspace{0.5cm}
\end{figure}

If we take the central value of $R_b$ reported at LEP and assume that
the discrepancy with the standard model value is due to SUSY, the
superpartners must be {\it quite} light.  This has recently been
analyzed in detail by Wells, Kane, and Kolda \cite{kolda}. For
$\tan\beta < 30$, the bounds on the chargino and top squark masses are
shown in Fig. \ref{graph1}.  They conclude that (1) if the reduction
in $R_b$ is due to SUSY, superpartners must be discovered either at
LEP II or the Tevatron and (2) the mass spectrum required cannot be
accommodated in the popular ``constrained'' minimal grand-unified
supersymmetry scenarios.

\begin{figure}[htb]
\begin{center}
\begin{tabular}{cc}
\epsfxsize 4cm \epsffile{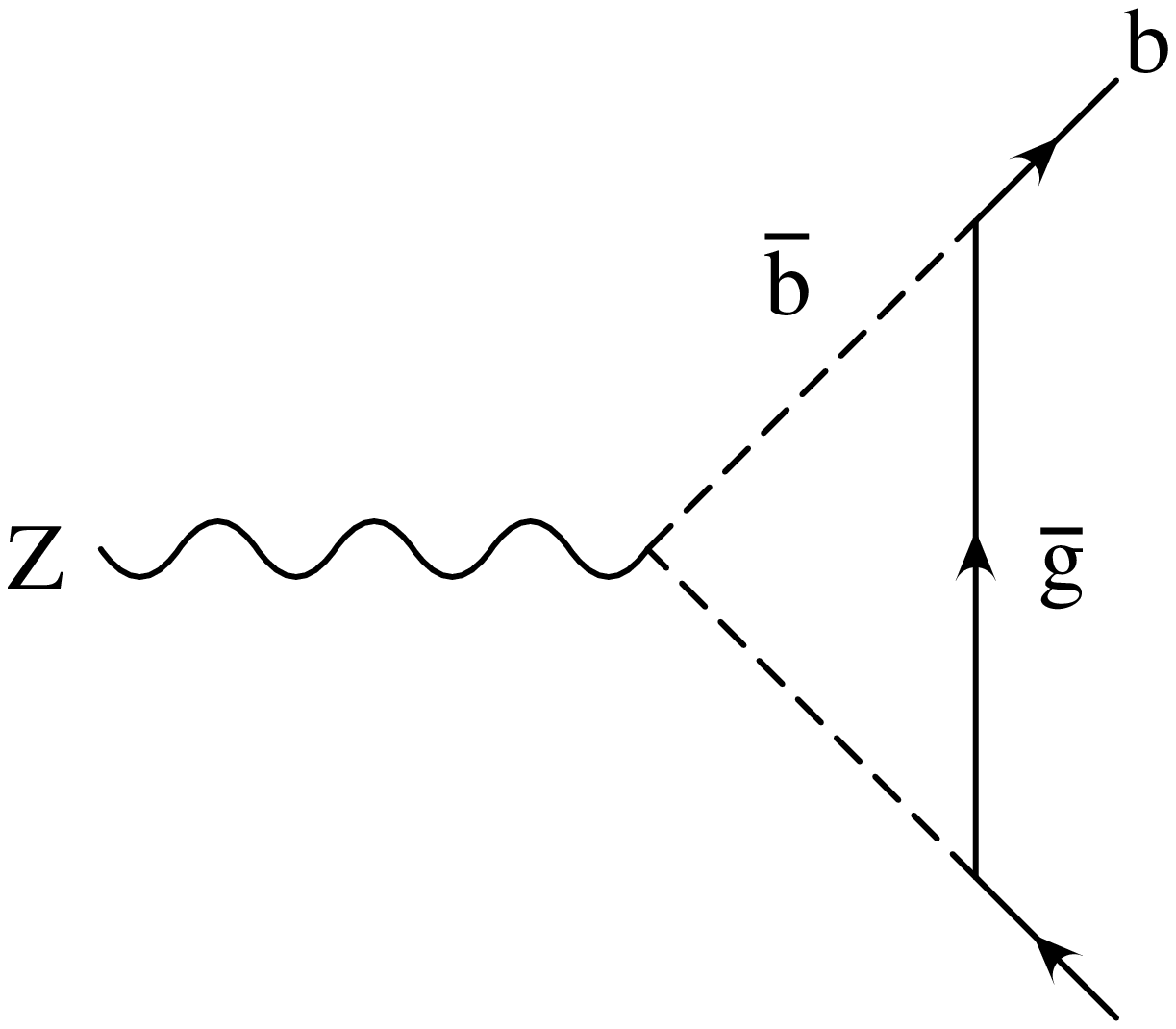}&
\epsfxsize 4cm \epsffile{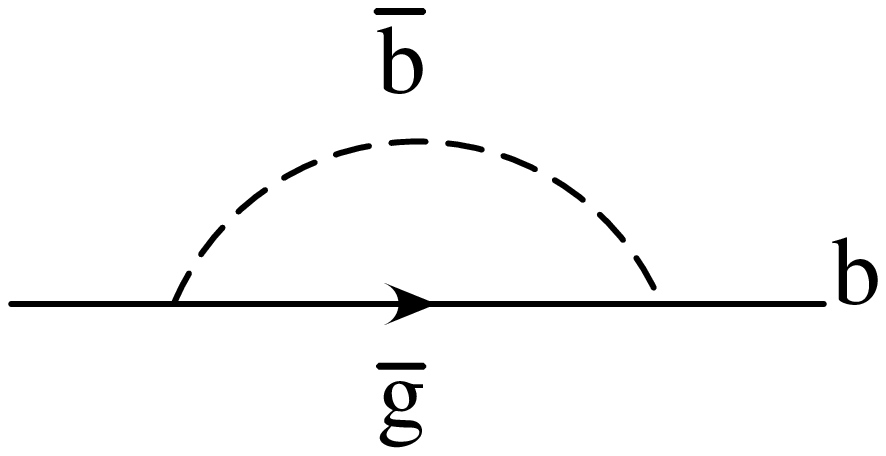}
\end{tabular}
\end{center}
\caption{ Potential strong SUSY corrections to $R_b$.}
\label{gluino}
\end{figure}

Finally, we should note that there are other new contributions to $R_b$ in
SUSY, including even some strong corrections involving the gluino, as
shown in Fig. \ref{gluino}. These have recently been calculated by Bhattacharya
and Raychaudhuri \cite
{bhatta}; however they are very small: the contributions are entirely
decoupling (they are not proportional to $m_t$) and vanish in the limit
that there is no $\widetilde{b}_L\leftrightarrow \widetilde{b}_R$
mixing, which is the only
$SU(2)\times U(1)$ breaking contribution to this process.

\section{Technicolor}

We move now to a completely different sort of theory, one with dynamical
electroweak symmetry breaking. In these theories, the electroweak symmetry
is broken due to the vacuum expectation value of a fermion bilinear instead
of that of a fundamental scalar particle

\be
\langle \varphi \rangle \to
\langle \bar \psi _L\psi _R\rangle .
\ee

\noindent
In the simplest theory \cite{technicolor} one introduces doublet of new
massless fermions

\be
T_L=\left(
\begin{array}{c}
U\\D
\end{array}
\right) _L\,\,\,\,\,\,\,\,
U_R,D_R
\ee

\noindent
which are $N$'s of an (asymptotically-free) technicolor gauge group
$SU(N)_{TC}$. In the absence of electroweak interactions, the Lagrangian for
this theory may be written

\bea
{\cal L} &=& \bar{U}_L i\slashchar{D} U_L+
\bar{U}_R i\slashchar{D} U_R+\\
 & &\bar{D}_L i\slashchar{D} D_L+
\bar{D}_R i\slashchar{D} D_R
\eea

\noindent
and thus has an $SU(2)_L\times SU(2)_R$ chiral symmetry. In analogy with QCD,
we
expect that when technicolor becomes strong,

\be
\langle \bar U_LU_R\rangle
=\langle \bar D_LD_R\rangle \neq 0,
\ee

\noindent
which breaks the global chiral symmetry group down to $SU(2)_{L+R}$, the
vector subgroup (analogous to isospin in QCD).

If we weakly gauge $SU(2) \times U(1)$, with the left-handed technifermions
forming a weak doublet and identify hypercharge with a symmetry generated by
a linear combination of the $T_3$ in $SU(2)_R$ and technifermion number,
then chiral symmetry breaking will result in the electroweak gauge group's
breaking down to electromagnetism. The Higgs mechanism then
produces the appropriate masses for the $W$ and $Z$ bosons if the $F$%
-constant of the technicolor theory (the analog of $f_\pi$ in QCD) is
approximately 246 GeV. (The residual $SU(2)_{L+R}$ symmetry insures
that the weak interaction $\rho$-parameter equals one at
tree-level \cite{weins}.)

\begin{figure}[htb]
\vspace{0.5cm}
\epsfxsize 6cm \centerline{\epsffile{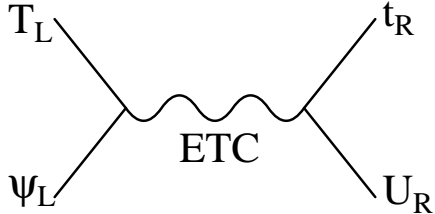}}
\caption{ ETC gauge-boson responsible for t-quark mass.}
\label{siete}\vspace{0.5cm}
\end{figure}

While this mechanism works wonderfully for breaking the electroweak symmetry
and giving rise to masses for the $W$ and $Z$ bosons, it does not account
for the non-zero masses of the ordinary fermions. In order to do so, one
generally introduces additional gauge interactions, conventionally called
``extended technicolor'' (ETC) interactions \cite{eicht},
which couple the chiral
symmetries of the technifermions to those of the ordinary
fermions (see Fig. \ref{siete}).

At low energies, below the mass of the ETC gauge boson, these interactions
may be approximated by local four-fermion interactions, and include a
coupling  of the following form

\be
\frac{g_{ETC}^2}{M_{ETC}^2}%
( \overline{T}_LU_R)(\overline{t}_R\psi _L).
\ee

\noindent
After technicolor chiral symmetry breaking, this interaction leads to a
mass for a quark (in this case the top-quark) of order

\be
m_t\cong \frac{g_{ETC}^2}{
\begin{array}{c}
M_{ETC}^{2_{}} \\
\end{array}
}\langle \bar U_LU_R\rangle .
\ee

It is the introduction of these extended technicolor interactions that
is the source of many of the problems of theories with dynamical
electroweak symmetry breaking. For example, in an ordinary QCD-like
technicolor theory it is difficult to arrange for the strange-quark
mass without introducing unacceptably large flavor-changing
neutral-currents. There are various ways around this and other
difficulties of ETC theories, for a review see \cite{lane}.

\begin{figure}[htb]
\vspace{0.5cm}
\epsfxsize 6cm \centerline{\epsffile{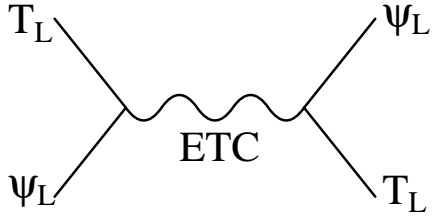}}
\caption{ETC interactions which gives rise to correction to
$R_b$.}
\label{ocho}\vspace{0.5cm}
\end{figure}

The ETC interactions produce corrections to
the $Zb\bar b$
branching ratio. The ETC gauge boson pictured in
Fig. \ref{siete} also mediates the
interaction shown in Fig. \ref{ocho}.

\begin{figure}[htb]
\vspace{0.5cm}
\epsfxsize 6cm \centerline{\epsffile{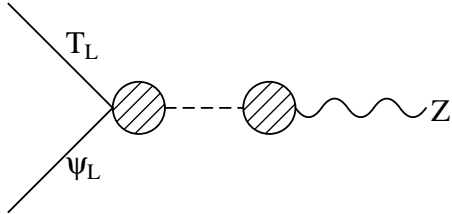}}
\caption{ ETC correction to $R_b$.}
\label{nueve}\vspace{0.5cm}
\end{figure}

At energies below the ETC gauge-boson mass, this interaction includes a
coupling that can be
approximated as

\be
\xi^2\frac{g_{ETC}^2}{%
M_{ETC}^2}\left( \overline{T}_L\gamma ^\mu \frac{\overrightarrow{\tau }}%
2T_L\right) \left( \overline{\psi }_L\gamma _\mu \frac{\overrightarrow{\tau }%
}2\psi _L\right)
\ee

\noindent where $\xi$ is a model-dependent Clebsch-Gordon coefficient
equal to one in the simplest models.  At energies below the
technicolor chiral symmetry breaking scale, this gives rise to the
interaction shown in Fig. \ref{nueve} and results in a change in
$g_L^b$ \cite{rsc}. Assuming for the moment that technicolor is
QCD-like, we can estimate the size of this effect and find

\be
\frac{\delta \Gamma }\Gamma
=-6.5\;\%\;\xi ^2\left( \frac{m_t}{175{\rm GeV}}\right)
\ee

\noindent
 From this we find $\delta R_b^{ETC}=-0.011\xi ^2$, which results in
a total $R_b\approx 0.205$ (for $\xi=1$).
This is approximately eight-$\sigma $ from the
reported value.

Of course, ordinary technicolor theories were already in trouble for
the flavor-changing neutral-current problems that were mentioned
previously. Unfortunately, even in the more popular walking
technicolor theories (so-called because the technicolor coupling runs
slowly at energies above the technicolor scale, and capable of
accommodating reasonable $s$ and $c$ quark masses without unreasonably
large flavor-changing neutral-currents) the effect is perhaps a factor
of two smaller \cite{gates} and is still hard to reconcile with
experiment.

As with other new physics, TC/ETC theories give contributions to $R_b$
which do not decouple with $m_t$. Furthermore, unlike the theories
discussed previously, we {\it cannot} take $M_{ETC}/g_{ETC}$ to be
arbitrarily high\footnote{Unless we include additional low-energy
scalar degrees of freedom \cite{ref:setc,ref:ccl}.}: its value is set by the
mass
of the top quark. In this sense, the contributions from TC/ETC
theories are {\it completely} non-decoupling -- their scale is set by
the masses of the gauge bosons or quarks. It is this fact which makes
the construction of phenomenologically acceptable TC/ETC theories so
difficult.

In the discussion above we have implicitly assumed that the gauge
bosons of the ETC theory do not carry electroweak quantum
numbers. Recently, we have begun \cite{liz} to investigate the
properties of theories containing ETC bosons which carry weak
charge. In this case, it is possible for the correction to be the same
order of magnitude, but {\it positive}. Such a correction may be too
large in the opposite direction (it would be off by four-$\sigma $) --
however such theories also include extra $Z$-bosons with
flavor-dependent couplings.  As we argue in the next section, these
extra effects may possibly bring TC/ETC theories into a
phenomenologically acceptable range.

\section{Extra Gauge Bosons}

The last class of physics beyond the standard model which I will discuss
concerns theories with extra weak gauge bosons. For simplicity, let us
consider theories with an extra $U(1)$ gauge symmetry, resulting in an extra
gauge boson $X$ which will mix with the ordinary $Z$. Following
the notation of Holdom \cite{holdon}, the terms in the Lagrangian
responsible for mixing include

\be
{\frac 12}M_Z^2Z^\mu Z_\mu +%
{\frac 12}M_X^2X^\mu X_\mu +xM_Z^2X^\mu Z_\mu ,
\ee

\noindent
where we have chosen a basis so that the field $Z_\mu $ has the conventional
gauge-couplings to the ordinary fermions and where we have neglected
additional kinetic-energy mixing terms (which are small in weakly-coupled
theories). In the limit $M_Z^2\ll M_X^2$, this mixing results in a change in
the coupling of the light mass-eigenstate (which we identify as the $Z$)

\be
\delta g_{L,R}^f\approx -x{%
\ \frac{M_Z^2}{M_X^2}}g_{X_{L,R}}^f.
\ee

The mixing, therefore, results in a change in the width of the $Z$ to
various fermions (including, in particular, the $b$). In addition it also
results in potentially dangerous changes in the relationship between $\sin
{}^2\theta _W$ to $\alpha $, $G_F$, and $M_Z$. For this reason, care must be
taken in extracting limits on extra gauge bosons from precisely measured
electroweak quantities \cite{altarelli}.

In ordinary extra gauge-boson models, of a type ``inspired'' by superstrings
or $SO(10)$ GUT models, the $X$ is usually assumed to couple to up-
and down-quarks in a flavor universal fashion. In the limit $M_X\to \infty $,
the theory reduces to the standard model. In this case, constraints on $
\delta R_b$ give constraints on $M_X^2$ and $g_X$.

In ETC/TC inspired models, however, the $X$ can be related to the gauge
boson responsible for generating the top-quark mass. For example, the
$X$ may be
a ``diagonal'' generator associated with ETC breaking at
the scale of top-quark mass generation. Therefore, in such theories it is
natural to assume that such a gauge boson couples more strongly to the
$b_L$,
$t_L$, and $t_R$ (and perhaps, though more dangerously, also to $b_R$, $
\tau _{L,R}$, and $\nu _\tau $). In such theories it is not possible to take
$M_X\to \infty $, since the mass of the $X$ is related to the size of $
m_t$ -- in this sense, the contributions are again completely non-decoupling.

The effects of such an extra family-dependent gauge boson are model
dependent. In theories where the ETC gauge-boson responsible for generating
the top-quark mass carries electroweak quantum numbers, the extra gauge
bosons result in a {\it decrease} to $R_b$ -- perhaps by an amount
sufficient to reconcile the ETC theory with experimental results. In a
four-generation ETC model introduced by Holdom \cite{holdon2},
the theory does {\em not} give
rise to an ETC contribution of the type discussed in the previous section
but an extra weak-singlet $X$ boson can {\it increase} $R_b$. These
theories need to be studied in greater detail, perhaps by the time of DPF
`96 we will understand them more fully and be able to determine whether they
can be consistent with the experimentally measured value of $R_b$.

\section{Perspectives}

Although we have discussed only one process in detail, there is a general
point that applies to the effects of physics beyond the standard model to
any precisely measured electroweak quantity. Namely, that there are two
types of theories of physics beyond the standard model.

First, there are {\em decoupling theories} that reduce to the standard
model in the limit where some parameter with the dimensions of mass is taken
to infinity. Generally, theories of this type investigated in the literature
are {\it weakly coupled. }Examples include:

\bigskip

$\bullet$ Two-Higgs Doublet Models (decouple when $m_{H^+} \to \infty$)

$\bullet $ Supersymmetric Theories (decouple when $m_{\widetilde{H}},
m_{\widetilde{q}} \to\infty $)

$\bullet$ Extra Gauge-Boson Theories, if it is possible for $M_X \to
\infty$

\bigskip

\noindent
Theories of this sort have both good and bad points. On the one hand,
because they reduce to the standard model in the decoupling limit, these
theories {\em cannot} be ruled out on the basis of precision electroweak
measurements (at least to the extent that these measurements are consistent
with the standard model). On the other hand, it is disappointing that the
answers to many of the interesting questions (SUSY breaking, Higgs masses,
the origin of flavor, etc.) may be hidden at {\it very} high ($M_{GUT}$ or
$M_{Pl}$) energy scales. In his talk, Jeff Harvey \cite{harvey} put the best
face on this issue by arguing that there may be enough clues in low-energy
parameters (e.g. the superpartner spectrum) to infer the
properties of the
high-energy physics. However, it is also possible that there will not be
enough clues at low-energies to shed light on the high-energy physics.

Second, there are {\em non-decoupling theories}, whose scales are fixed by
the masses of the observed particles. Theories of this type
generally discussed in the literature are {\it strongly} coupled, and this
makes them somewhat difficult to analyze. Examples include

\bigskip

$\bullet$ Technicolor/ETC Theories (the technicolor scale, $\Lambda_{TC}$
-- the analog of $\Lambda_{QCD}$ in the ordinary strong interactions, is
fixed at the weak scale).

$\bullet$ Extra Gauge-Boson Theories, in which $M_X$ is fixed by the
top-quark or some other mass.

\bigskip

\noindent
If one succeeds in constructing a theory of this sort, one has made an
enormous amount of progress -- such a theory would explain a lot with
physics at accessible energies (of order a TeV). However, there are no fully
realistic models of this sort. Generically, these theories predict
{\it large} low-energy effects of a sort that are excluded experimentally.

With luck, in time we will have experimental evidence to decide which type
of theory is operative in the real world. However, given how little we
actually know about the dynamics of electroweak and flavor symmetry
breaking, we should be ready for either possibility.

\section{Prospects}

In this talk I have concentrated on corrections to the coupling of
the $Z$
to $b$-quarks. This coupling is particularly susceptible to corrections
because the left-handed $b$, being in the same weak doublet as the
left-handed $t$, couples to the physics responsible for generating the large
$t$-quark mass. However, one would like to probe the couplings of the
t-quark {\it directly}.

One possibility is that the physics of electroweak symmetry breaking and
flavor {\it could} lead to an enhancement of the cross section $\sigma
(pp\to t\bar t+X)$ at the Tevatron. Two proposals of this sort have been put
forward recently. In the first, due to Hill and Parke
\cite{hill}, there is
an additional color-octet or singlet gauge-boson, perhaps from
``top-color'' interactions \cite{top}, which is produced in $q\bar q$
annihilation. In the second, due to Eichten and Lane\cite{eich-lane},
a color-octet pseudo-Goldstone boson (a colored analog of the $\eta $
in QCD, expected in some technicolor models) is produced in gluon
fusion. In both cases, the new particle is associated with electroweak
symmetry breaking and therefore couples most
strongly and decays preferentially to the top quark.

These scenarios are particularly interesting because the rate of top
quark production, $\sigma =13.9_{-4.8}^{+6.1}$ pb (assuming that the
excess of leptons plus jet events observed at the Tevatron is due to a
174 GeV top quark), is somewhat higher than the theoretically
predicted value, $\sigma^{t\bar{t}} _{QCD}=5.10_{-0.43}^{+0.73}$ pb
\cite{cdf}. With only a handful of events, we cannot be sure that the
top-quark production cross section is, in fact, higher than expected
from QCD. However, these models (1) demonstrate the often neglected
possibility that the electroweak symmetry breaking sector couples to
QCD and, more important, (2) will be tested in the near future as more
data is collected at the Tevatron.  These issues are discussed further
by Steve Parke \cite{parke} in these
proceedings.

A second possibility is that one may directly probe the couplings of the
top-quark to the $W$ and $Z$ gauge bosons. A preliminary analysis
by Barklow and Schmidt shows that it may be possible at an $e^{+}e^{-}$
collider, with $\sqrt{s}=500$ GeV and an integrated luminosity of
50 fb$^{-1}$, to measure these
couplings to an accuracy of 5\%-10\%. Details of this
work may be found in the contribution by Schmidt \cite{schmidt}
in these proceedings.

\section{Summary}

In conclusion let me reiterate that, because the top quark is heavy, it
couples more strongly to the symmetry breaking sector. In general, this may
be viewed as due to the  Goldberger-Treiman relation, eqn.
\ref{gt}. Therefore, the top quark may provide a window on {\it both}
electroweak and flavor symmetry breaking. Furthermore, because of $SU(2)_W$
symmetry, the physics responsible for generating the large $t$-quark mass
also couples to the left-handed component of the $b$, resulting in
contributions to the $Zb\bar b$ branching ratio which are generically
{\it non-decoupling} in $m_t$ (and therefore enhanced).

Will the measured value of $R_b$ remain above the standard model
value? Only time will tell.

\section{Acknowledgements}

I thank Mike Dugan, Kenneth Lane, Elizabeth Simmons, and John Terning for
discussions and for comments on the manuscript and gratefully
acknowledge the support of an Alfred P. Sloan Foundation Fellowship,
an NSF Presidential Young Investigator Award, and a DOE Outstanding
Junior Investigator Award. This work was supported in part under NSF
contract PHY-9218167 and DOE contract DE-FG02-91ER40676.

\newpage
\renewcommand{\Large}{\normalsize} 

\end{document}